\documentclass[12pt]{spieman}  
\usepackage{amsmath,amsfonts,amssymb}
\usepackage{graphicx}
\usepackage{setspace}
\usepackage{tocloft}
\usepackage{xspace}
\usepackage{bm}
\usepackage{lineno}

\newcommand{\HST}{\textit{HST}\xspace}
\newcommand{\JWST}{\textit{JWST}\xspace}
\newcommand{\IRSSquare}{IRS$^2$\xspace}
\newcommand{\WFIRST}{\textit{WFIRST}\xspace}

\newcommand{\bnm}[1]{\bm{\mathbf{#1}}}
\newcommand{\ie}{{\it i.e.}\xspace}

\newcommand{\etc}{{\it etc.}\xspace}
\newcommand{\etal}{{\it et al.}\xspace}

\newcommand{\arcdeg}{\ensuremath{^{\circ}}}

\title{Principal Component Analysis\\ of Up-the-ramp Sampled IR Array Data\footnote{Accepted by the Journal of Astronomical Telescopes, Instruments, and Systems (JATIS). Please reference this work as Rauscher, B.J. et al. (2019), JATIS, in press.}}

\author[a,*]{Bernard J. Rauscher}
\author[b]{Richard G. Arendt}
\author[c]{\\D.J. Fixsen}
\author[c]{Alexander Kutyrev}
\author[a]{Gregory Mosby}
\author[a]{S.H. Moseley}
\affil[a]{NASA Goddard Space Flight Center, Observational Cosmology Laboratory, Greenbelt, MD, USA, 20771}
\affil[b]{CRESST II/UMBC/GSFC,Greenbelt, MD}
\affil[c]{CRESST/UMd/GSFC, Greenbelt, MD}

\cftpagenumbersoff{figure}
\cftpagenumbersoff{table} 
\begin{document} 
\maketitle

\begin{abstract}
We describe the results of principal component analysis (PCA) of up-the-ramp sampled IR array data from the \HST WFC3 IR, \JWST NIRSpec, and prototype \WFIRST WFI detectors.  These systems use respectively Teledyne H1R, H2RG, and H4RG-10 near-IR detector arrays with a variety of IR array controllers. The PCA shows that the Legendre polynomials approximate the principal components of these systems (\ie they roughly diagonalize the covariance matrix). In contrast to the monomial basis that is widely used for polynomial fitting and linearization today, the Legendre polynomials are an orthonormal basis. They provide a quantifiable, compact,  and (nearly) linearly uncorrelated representation of the information content of the data. By fitting a few Legendre polynomials, nearly all of the meaningful information in representative WFC3 astronomical datacubes can be condensed from 15 up-the-ramp samples down to 6  compressible Legendre coefficients per pixel. The higher order coefficients contain time domain information that is lost when one projects up-the-ramp sampled datacubes onto 2-dimensional images by fitting a straight line, even if the data are linearized before fitting the line. Going forward, we believe that this time domain information is potentially important for disentangling the various non-linearities that can affect IR array observations, \ie inherent pixel non-linearity, persistence, burn in, brighter-fatter effect, (potentially) non-linear inter-pixel capacitance (IPC), and perhaps others.
\end{abstract}


{\noindent \footnotesize\textbf{*}Bernard J. Rauscher,  \linkable{Bernard.J.Rauscher@nasa.gov} }

\begin{spacing}{1}   

\section{Introduction}\label{sec-intro}

The Wide Field Infrared Survey Telescope's (\WFIRST)\cite{2015arXiv150303757S} cosmology and microlensing surveys require excellent control of detector systematics. For example, the \WFIRST High Galactic Latitude Weak Lensing Survey requires knowledge of the size and ellipticity of the point-spread function (PSF) to $<0.1\%$ to avoid biasing measurements of dark energy properties and other cosmological parameters.\cite{Shapiro:2013gx} However, the PSF's shape is measured using reference stars that are typically much brighter than weakly lensed field galaxies, necessitating large linearity corrections.

When large linearity corrections are required, non-ideal instrument signatures including inherent pixel non-linearity, the brighter-fatter effect (caused by charge integrating in neighboring pixels as pixels fill up),\cite{Plazas:2018bw} residual persistence from prior exposures, (potentially) non-linear inter-pixel capacitance (IPC),\cite{Moore:2006kz} and telescope pointing jitter (this list is almost certainly incomplete) have the potential to compromise PSF knowledge and thereby \WFIRST science. Within the \WFIRST IR Detector Working Group, understanding non-linear pixel response is a particularly high priority. We therefore began our study by trying to better understand modern linearity correction techniques and the linearity properties of prototype \WFIRST IR arrays.

Today's IR calibration pipelines for missions including the Hubble Space Telescope's (\HST) Wide Field Camera 3 IR (WFC3)\cite{MacKenty:2008kj,MacKenty:2010jd,MacKenty:2012ky,Dressel:2018} and the James Webb Space Telescope (\JWST)\cite{Gardner:2006di} linearize the up-the-ramp sampled data from each pixel before fitting a straight line to infer brightness from the fitted slope.\footnote{For a description of up-the-ramp sampling, please see Appendix~\ref{sec-detector-info}. The WFC3 IR calibration pipeline is described in $\S 3.3$ of The WFC3 Data Handbook.\cite{Gennaro:2018} A similar pipeline is in development for \JWST and is planned for \WFIRST.} The planned pipeline calibration sequence for \JWST is typical. It includes: (1) bias correction, (2) reference pixel correction, (3) linearization, (4) dark subtraction, (5) cosmic ray detection, and (6) slope fitting. Linearization is based on fitting a low degree polynomial to the up-the-ramp samples in calibration flats. As described by Vacca, Cushing, and Raynor (2004)\cite{Vacca:2004fg} and Hilbert (2014)\cite{Hilbert:2014uy}, the resulting polynomial fit coefficients are used to linearize astronomical exposures before fitting a straight line to each pixel to infer the flux. Typically, linearization assumes that the charge integration rate at the beginning of the exposure is the ``true'' one, and uses the calibration polynomial fit coefficients to make multiplicative corrections to later samples for which non-linearity is significant.

Fitting a ``deg'' degree polynomial to $n$ up-the-ramp samples projects the data from the Cartesian $n$-space in which they were acquired into a monomial space of deg+1 dimensions, (the monomial basis vectors are $\{z^0,z^1,z^2,\ldots,z^{\rm deg}\}$, where $z$ is the time index for equally spaced samples). The coordinates in monomial space, the polynomial fit coefficients, provide a representation of a pixel's response that is optimal in a least squares sense to the specified fit degree. However, the monomials are not an orthogonal basis, and they do not offer insight into how high the fit degree needs to be. Moreover, as we  show in $\S \ref{sec-advantages}$, there often exist significant correlations between the monomial fit coefficients. For these reasons and others, we decided to use principal component analysis (PCA) to see if there might exist a better basis for modeling up-the-ramp sampled IR array data than the monomials.

The input data were provided by the NASA Goddard Space Flight Center Detector Characterization Laboratory (DCL) as part of the \WFIRST project. \WFIRST's Wide Field Instrument (WFI) uses 18 Teledyne H4RG-10 near-IR detector arrays. The 4K$\times$4K~pixel H4RG-10\cite{Piquette:2014cc} is the most recent member of Teledyne's HxRG family of HgCdTe near-infrared detector arrays. The underlying HxRG architecture is an outgrowth of the 2K$\times$2K~pixel H2RG\cite{Loose:2003vh} that was introduced in about 2003 for \JWST. The ``H'' in HxRG stands for Hawaii, $x\in\{1,2,4\}$ refers to the number of kilopixels in the vertical and horizontal directions, ``R'' indicates the presence of reference pixels, and ``G'' indicates the availability of guide mode. Compared to the earlier 1K$\times$1K~pixel H1R detector that was used by WFC3, the H2RG and H4RG include a built-in guide mode. Two versions of the H4RG are available. The H4RG-10 has 10~$\mu$m pixels and is used by \WFIRST because mass and volume are at a premium. The physically larger H4RG-15\cite{Zandian:2016cq} offers 15~$\mu$m pixels. It may be better matched to the optics of large, ground based telescopes.

At first, we studied laboratory flatfield data from two different \WFIRST H4RG-10s controlled by $3^{\rm rd}$ generation ``Leach'' controllers from Astronomical Research Cameras, Inc., and both gave identical results. The PCA immediately suggested a dramatically better basis than the monomials: the Legendre polynomials. If $z$ is an index that runs over frame number and $s(z)$ is integrated signal, then today's monomial approach fits,
\begin{equation}
s(z)=\sum_{i=0}^{\rm deg} a_i z^i,
\end{equation}
where the $a_i$ terms are the fit coefficients. Typically $\rm deg=3$ is used for linearization while $\rm deg = 1$ is used for slope fitting (see Ref.~\citenum{Hilbert:2014uy} and references therein). When fitting Legendre polynomials,  $z$ is uniformly mapped to the interval $-1\leq x\leq +1$, and signal integrates according to,
\begin{equation}
s(x)=\sum_{i=0}^{\rm deg} \lambda_i P_i\left(x\right).
\end{equation}
In this expression, $\lambda_i$ are the fit coefficients and the Legendre basis vectors are\\ $\left\{P_0\left(x\right),P_1\left(x\right),P_2\left(x\right),\ldots, P_\text{deg} \right\}$.

Unlike the monomials, the Legendre polynomials are an orthonormal basis. Moreover, because the Legendre polynomials approximately diagonalize the covariance matrix, they provide a representation that is significantly less correlated than the monomials. Finally, because the covariance matrix is roughly diagonal in Legendre space, the Legendre fit coefficients provide a useful way to quantify how the information content falls off by fit degree.

Although we began the PCA using \WFIRST H4RG-10 laboratory flats, we have since shown that the  results are general; applying equally to other systems including the \JWST Near Infrared Spectrograph and \HST WFC3. The findings do not depend on the type of IR detector or the IR array controller (H1R+Ball electronics; H2RG+SIDECAR ASIC; H4RG-10 + Gen. III Leach controller). All systems have given similar results. These are the findings.

\begin{enumerate} 
\item The Legendre polynomials generally provide a good basis for the time dimension of up-the-ramp sampled IR array data. The monomial basis that is used in today's pipelines is a comparatively poor choice because the basis vectors are not orthogonal and monomial fit coefficients are correlated.
\item Up-the-ramp sampled IR array data contain time domain information that is not utilized by fitting a straight line to the linearized up-the-ramp samples. As a corollary, for archival research it is important to save more information than is contained in only the fitted bias and slope images.
\item Legendre fits are highly compressible. If it is not possible or practical to save all samples, then Legendre fitting offers a simple way to significantly reduce the data volume while retaining nearly all of the scientific information.
\end{enumerate}

Legendre fitting naturally produces datacubes rather than 2-dimensional (2D) images. Two kinds of datacube are commonly encountered in near-IR astronomy today. Often, the unprocessed up-the-ramp samples are saved in a 3D array, with the axes being time (or equivalently time index for equally space samples) and the two angular dimensions on the sky, $\left(t,\alpha,\delta\right)$. Another common astronomy datacube has wavelength running along the zeroth dimension. Here we introduce the idea of a ``Legendre cube''. A Legendre cube differs from a time-ordered 3D array in that the zeroth dimension contains the fit coefficients, $\lambda_i$, for the Legendre polynomial $P_i(x)$. The axes of a Legendre cube are $\left(i,\alpha,\delta\right)$.

This article is structured as follows. In $\S \ref{sec-pca}$, we explain the PCA in the context of WFC3's Frontier Field observations of the Abell 370 strong lensing field. We selected Abell 370 because the field is information rich, and it contains a wide range of source brightness ranging from sky background to hard saturation. This is followed by $\S \ref{sec-info}$, where we explore how  astrophysical information manifests in Legendre cubes. The higher Legendre orders are information poor, leading to $\S \ref{sec-compress}$, where we discuss data compression in Legendre space. Finally, in $\S \ref{sec-astro}$, we begin to describe how these findings can potentially be applied to astrophysical observations. This section is necessarily incomplete as we are only beginning to work with Legendre cubes today. We close with a summary.

\section{Principal Components Analysis}\label{sec-pca}

The purpose of this section is to describe the PCA in the context of real WFC3 observations. We previously obtained similar results with \WFIRST and \JWST lab data (including both flat field images and un-illuminated darks). We begin with a short introduction to the WFC3 data. This is followed in $\S \ref{sec-refresh}$ by a PCA refresher, and then by the PCA itself.

\subsection{Abell 370 Frontier Field}\label{sec-ff}

\begin{figure}[h]
\begin{centering}
\includegraphics[width=85mm]{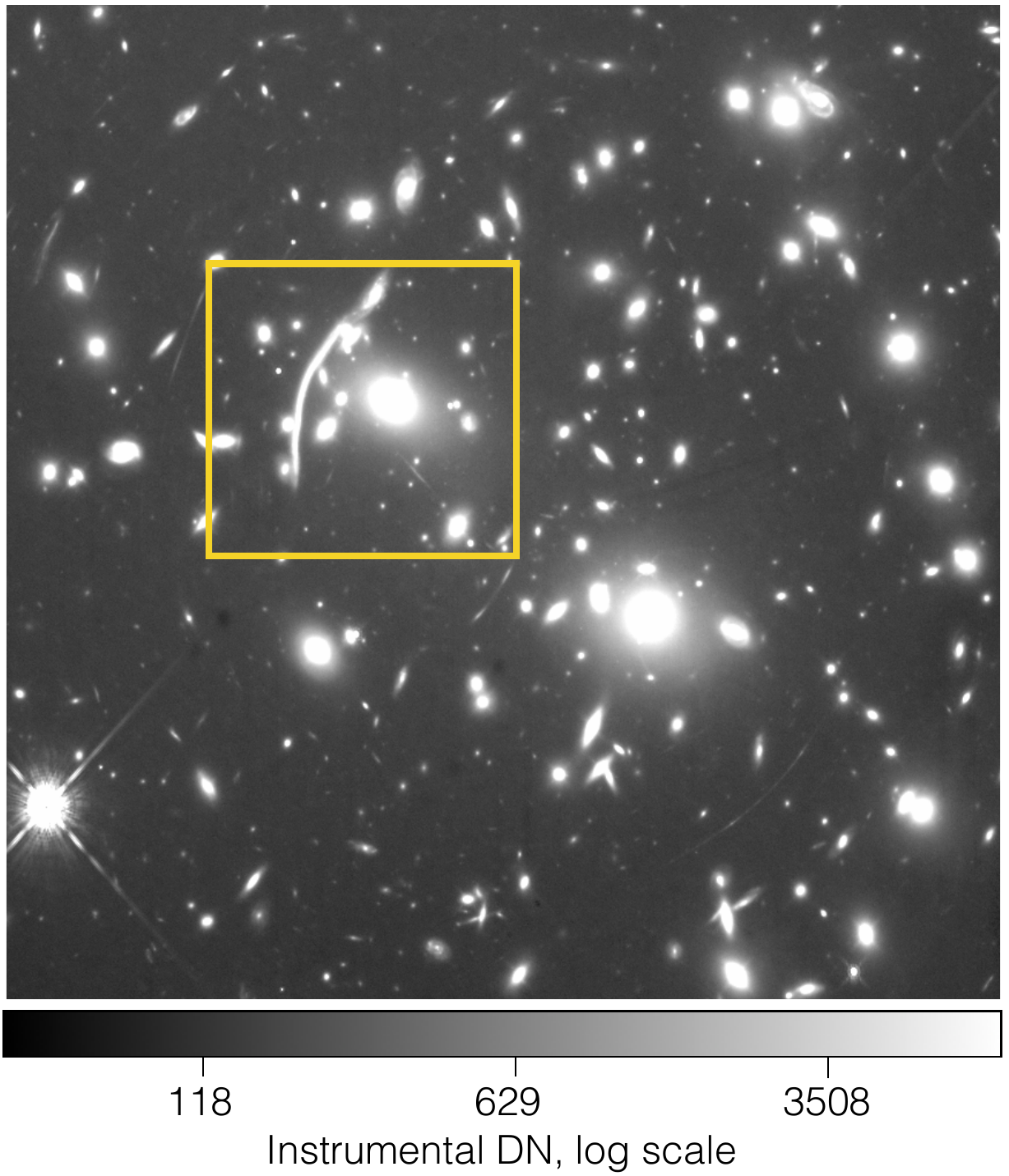}
\caption{This image is the median $\lambda_1$ parameter (slope) from stacking the available WFC3 F160W data. The integrated exposure time is about 14~hours. Other than reference correction (Appendix~\ref{sec-detector-info}) and fitting Legendre polynomials, the image is uncalibrated. The yellow box is the region of interest (ROI) shown in Figure~\ref{fig-lambdas}.}\label{fig-overview}
\end{centering}
\end{figure}

The Frontier Fields were an \HST Director's Discretionary program that aimed to exploit the amplification of light by strong gravitational lensing to image faint, high-$z$ galaxy populations.\cite{Lotz:2017ic} We selected one of the Frontier Fields, Abell 370 (Figure~\ref{fig-overview}), to test the ideas that we had conceived earlier using \WFIRST lab data.

The data were acquired between August and September, 2016, as part of proposal ID 14038 (J. Lotz PI). We selected all available  16~frame SPARS100 exposures taken with the F160W filter ($\lambda_\text{peak}=1.545~\mu$m, $\text{FWHM}=0.29~\mu$m).\footnote{We also looked at other filters. The choice of filter does not affect the PCA results.} In SPARS100 mode, WFC3 acquires a ``reset frame'' followed by 15 uniformly spaced non-destructive samples at $\sim 100~s$ intervals. Because the reset frame did not follow the same approximately linear trend as later samples, it was discarded (it is also not used for fitting in the WFC3 pipeline).  All told, we were able to use 36 $\text{EXPTIME}\approx 1403~s$ dithered SPARS100 exposures, each with 15 up-the-ramp reads, resulting in a total exposure time of about 14 hours.

Our focus in this paper is the PCA. For more information about WFC3 readout modes and data products, the interested reader is referred to The WFC3 Instrument Handbook\cite{Dressel:2018} and The WFC3 Data Handbook.\cite{Gennaro:2018}

\subsection{PCA Refresher}\label{sec-refresh}

Up-the-ramp sampled data are typically stored in datacubes, or in the case of WFC3 as FITS files with one extension per frame. For these WFC3 data, the unprocessed files unpacked to $15\times 1024\times 1024$ datacubes after discarding the reset frames. Although time-ordered datacubes are an intuitively obvious way to store  data, they are neither the most compact nor the most useful way to represent the information. PCA is a mathematical tool for finding a more compact representation that reduces the number of variables that are needed to represent the dataset while retaining much of the information.\cite{Jolliffe2011} Throughout this paper, we use lowercase boldface letters to represent vectors and uppercase boldface letters to represent matrices.

PCA is built on the concept of the covariance matrix, $\bnm{\Omega}$, which is a generalization of the variance that allows for the possibility of correlations. Consider a univariate statistical process, $\bnm{d}$.  For present purposes, $\bnm{d}$ is a vector containing 15 up-the-ramp samples. If the expectation value of $\bnm{d}$ is known to be $\langle \bnm{d}\rangle$, then the residuals for any one realization can be represented by a column vector,
\begin{equation}\label{eq-delta}
\bnm{\delta}=\bnm{d}-\langle \bnm{d}\rangle.
\end{equation}
If the experiment is run $n$ times (or we have $n$ pixels), we can put all of the different realizations of $\bnm{\delta}$ into a matrix, $\bnm{\Delta}$. The covariance matrix is then defined,
\begin{equation}\label{eq-cov}
\bnm{\Omega} = \frac{1}{n}\bnm{\Delta} \bnm{\Delta}^{\rm T}.
\end{equation}

$\bnm{\Omega}$ is by definition square, symmetric, and positive definite. If there is any correlation in the data (there is with up-the-ramp sampled detector data), then $\bnm{\Omega}$ will have off diagonal elements. In PCA, we seek a basis in which the covariance matrix is diagonal.

Since $\bnm{\Omega}$ is  square, by the eigen decomposition theorem it can be factored,
\begin{equation}
\bnm{\Omega}=\bnm{V}  \bnm{W}  \bnm{V}^{-1}.
\end{equation}
In this expression, $\bnm{W}$ is a diagonal matrix containing the positive eigenvalues of $\bnm{\Omega}$. $\bnm{V}$ is a square matrix containing the corresponding eigenvectors (one eigenvector per column). Because $\bnm{\Omega}$ is symmetric, the eigenvectors form an orthogonal basis. In the PCA lexicon, each orthogonal basis vector corresponds to a ``component''.

Because each column of $\bnm{V}$ is a basis vector, applying $\bnm{V}^{\rm T}$ to the data, $\bnm{d^\prime}=\bnm{V}^{\rm T}\bnm{d}$,  accomplishes a change of basis yielding, $\bnm{d^\prime}$, that contains the same information but with orthogonal deviates (\ie the covariance matrix is diagonal when computed in this new basis).

This a more convenient representation because each component can now be considered independently.  It can save computation (ignoring the computation to convert it into this form) since weighting can be done component by component instead of multiplying by a matrix. The data can also be stored more compactly.  This is because each component only needs to be stored with enough precision to encompass the noise of that component. Moreover, in many physical systems most of the information occurs in the first few components.

\subsection{PCA Method}\label{sec-method}

We used nearly all of the pixels shown in Figure~\ref{fig-overview} (about 98.5\%) for the PCA. The $\sim$1.5\% that were discarded were found by doing a simple reference correction and preliminary $5^{th}$ degree Legendre fit. We then rejected high rms pixels based on a histogram. In practice, this is very loose quality control. Nevertheless, the PCA gave consistent results in spite of there being cosmic ray hits and other artifacts in the data.

After discarding the reset frame, reference correction, cropping off reference pixels, let the data from the $i^{th}$ of $n\approx 1014^2$ pixels be represented by a column vector,
\begin{equation}
\bnm{d}_i=\left(
\begin{array}{c}
 d_{0\ i} \\
 \vdots  \\
 d_{14\ i} \\
\end{array}
\right).
\end{equation}
By flattening the two spatial dimensions on the sky, the data can be represented by a matrix,
\begin{equation}
\bnm{D}=\left(
\begin{array}{ccc}
 d_{0\ 0} & \cdots  & d_{0 n-1} \\
 \vdots  & \cdots  & \vdots  \\
 d_{14\ 0} & \cdots  & d_{14 n-1} \\
\end{array}
\right).
\end{equation}
Each column in $\bnm{D}$ contains the up-the-ramp samples from one pixel.

Define the matrix, $\langle \bnm{D}\rangle$, to be a matrix that has the same shape as $\bnm{D}$, but with every column equal to a column vector that is the expectation value of $\bnm{D}$ averaged over all columns. All columns of $\langle \bnm{D}\rangle$ are therefore identical. The covariance matrix is then,
\begin{equation}\label{eq-omega}
\bnm{\Omega} =\frac{1}{n} \left(\bnm{D}-\langle \bnm{D}\rangle \right) \left(\bnm{D}-\langle \bnm{D}\rangle \right)^\mathrm{T}.
\end{equation}

\begin{figure*}[t]
\begin{centering}
\includegraphics[width=\textwidth]{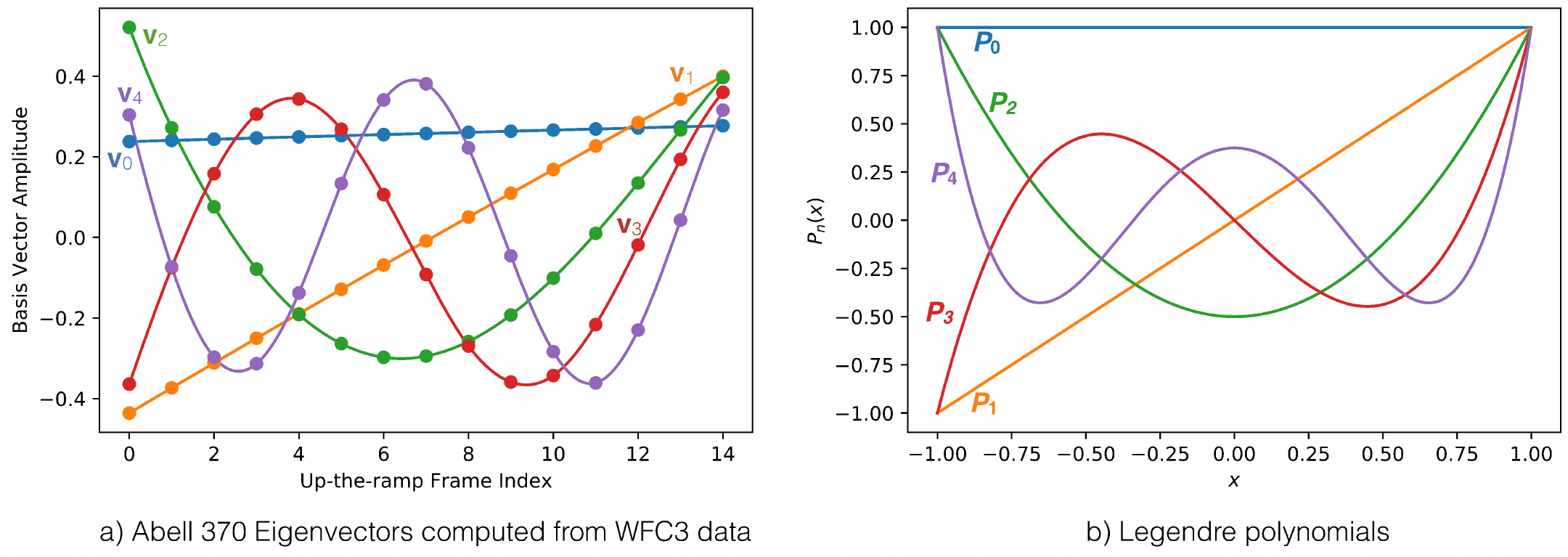}
\caption{Panel a) shows the first five eigenvectors computed from the WFC3 Abell 370 data. They are very similar to the b) Legendre polynomials. For comparison purposes, we show the actual Legendre polynomials in b). If we were to change the normalization to match a), then the two plots would look even more similar.}\label{fig-evecs-leg}
\end{centering}
\end{figure*}

To complete the PCA, following $\S \ref{sec-refresh}$ we computed  $\bnm{\Omega}$'s eigensystem. Figure~\ref{fig-evecs-leg}a shows the eigenvectors, $\bnm{v}$. By inspection, the eigenvectors are similar to the Legendre polynomials (Figure~\ref{fig-evecs-leg}b). For clarity, we show only the first few here, although the striking similarity continues as more are plotted. PCA of \WFIRST and \JWST flats (Appendix~\ref{PCA-of-flats}) produced similar results. However, we find this demonstration using real astronomical data to be particularly compelling. Moreover, the result is not particular to Abell~370. We did the same analysis for WFC3 observations of the 47~Tucanae globular cluster and got the same results.

\begin{figure*}[t]
\begin{centering}
\includegraphics[width=85mm]{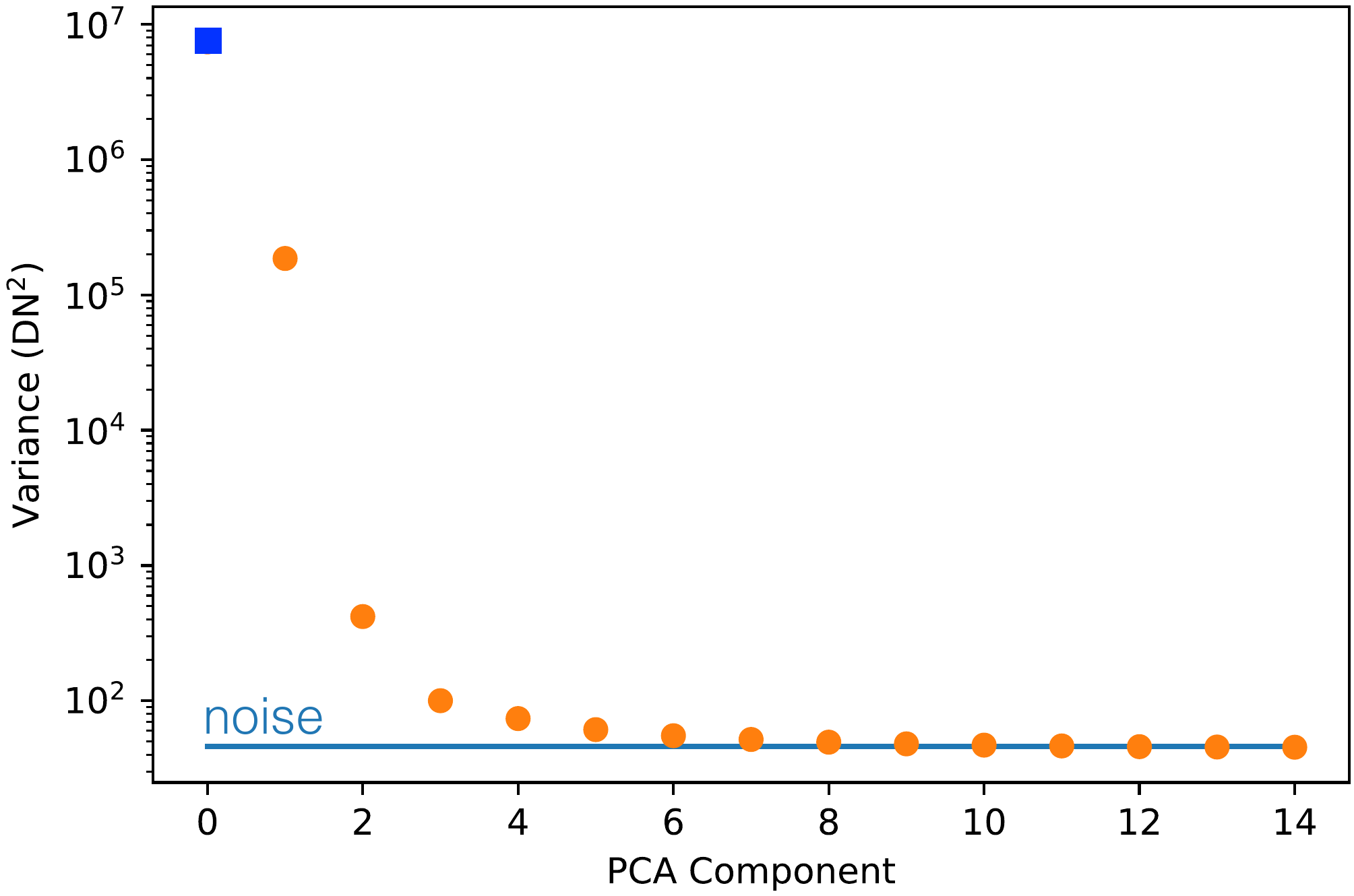}
\caption{This plot of eigenvalue vs PCA component ($\lambda$ index) shows that essentially all of the meaningful information is contained in $\lambda_0-\lambda_5$. The symbol for $\lambda_0$ is different because it is dominated by the instrument signature (detector bias pattern and hot pixels \etc). Excluding $>$3.5$\sigma$ statistical outliers, the noise is essentially Gauss-normal for $\lambda_6$ and higher (Appendix~\ref{sec-gaussian-noise}). For reference, WFC3's conversion gain is about, $g_c\sim 2.25~e^-~\rm DN^{-1}$. According to the WFC3 Instrument Handbook, the read noise is between $20.2-21.4~e^-$ per correlated double sample.\cite{Dressel:2018} The read noise per sample is therefore about $15~e^-$, which corresponds to the variance of the blue noise line that is overlaid on the plot.}\label{fig-evals}
\end{centering}
\end{figure*}

Figure~\ref{fig-evals} shows the eigenvalues, $w_i$, plotted by PCA component. They provide a quantitative measure of the variance in each component. Throughout this paper, we equate variance with ``information''. Not all information is scientifically useful because some is just noise. We define the ``meaningful information'' to be the remaining variance after subtracting off the noise. Consistent with this plot, a variety of statistical tests (Appendix~\ref{sec-gaussian-noise}) show that the noise is essentially Gauss-normal for $\lambda_6$ and higher once $>$3.5$\sigma$ outliers are excluded. Later, in $\S \ref{sec-info}$, we will discuss images (Figure~\ref{fig-lambdas}) of information content by PCA component. For these observations, the evidence is consistent with nearly all of the scientific information being contained in $\lambda_0-\lambda_5$.

\subsection{Why Legendre Polynomials?}\label{sec-why}

It is reasonable to ask, is there a physical explanation for why the Legendre polynomials emerge so clearly? If one views the problem of fitting the up-the-ramp samples as both a physics problem and as a linear algebra problem, then one can plausibly argue that the first few Legendre polynomials ought to be a good basis for modeling a pixel's response to light.

Whenever a pixel is reset, there is a statistical uncertainty in the number of charges on the pixel's capacitance equal to, $\sigma_\textrm{kTC}=q_e^{-1}\sqrt{k_B T C}$.\cite{Rieke:2007fn} In up-the-ramp sampled data, this ``kTC noise'' appears as a constant offset affecting all samples. In a linear model parameterized by up-the-ramp frame index, $z$, kTC noise appears as a constant times $z^0$. After the pixel is reset, by design, the ideal pixel starts to integrate charge in a highly linear manner. We therefore expect a term that looks like a constant times $z^1$. As charge integrates, we know from looking at data that most pixels gradually roll over and lose response in a manner that can be roughly approximated with a quadratic, \ie a constant times $z^2$, before steepening upon entering harder saturation as a constant times $z^3$ (and so on). We can use these observations as the starting point for a linear model of the pixel's response that includes the basis set, $B=\{z^0,z^1,z^2,z^3 \}$.

Without loss of generality, we can map $z$ to the interval $-1\leq x\leq +1$, and furthermore define an inner product for this interval,
\begin{equation}
\begin{centering}
\langle a,b\rangle =\int_{-1}^{+1} a\left(x \right) b \left(x \right) \, dx.\label{eq-inner-product}
\end{centering}
\end{equation}
If one uses Equation~\ref{eq-inner-product} to orthogonalize $B$, the result is the first four Legendre polynomials, $B^\prime=\{P_0\left(t \right),P_1\left(t \right),P_2\left(t \right),P_3\left(t \right) \}$.\cite{0471504475} One can extend this argument to higher degree, although the justification for the higher order terms becomes less physical the higher one goes.

Viewed in this way, the Legendre polynomials emerge as a natural orthogonal basis for pixels that have kTC noise, respond to light in a roughly linear matter at low signal levels, and for which the response gradually rolls off before steepening upon entering saturation. Empirically, it so happens that they also approximately diagonalize the covariance matrix.

\subsection{Advantages of Legendre Polynomials}\label{sec-advantages}

One advantage of the Legendre polynomials compared to the monomials is that they provide a less correlated representation of the data. To quantify this, we looked at the Pearson correlation matrix, $\bnm{P}$ (also known as {\em Pearson's r}; see Ref.~\citenum{Press:1992:NRC:148286}, $\S 14.5$). To compute $\bnm{P}$, define the matrix $\bnm{\Delta}^\prime$ that has the same shape as $\bnm{D}$. The $i^{th}$ column of $\bnm{\Delta}^\prime$ is equal to the column vector,
\begin{equation}
\bnm{\delta}^\prime _i=\left(
\begin{array}{c}
 \frac{d_{0 i}-\left\langle d_0\right\rangle }{std\ d_0} \\
 \vdots  \\
 \frac{d_{14 i}-\left\langle d_{14}\right\rangle }{std\ d_{14}} \\
\end{array}
\right),
\end{equation}
where $std$ is the usual standard deviation. With these definitions, the Pearson correlation matrix becomes,
\begin{equation}
\bnm{P}=\frac{1}{n}\bm{\Delta}^\prime  \bm{\Delta}^{\prime\rm T}.
\end{equation}

\begin{figure*}[t]
\begin{centering}
\includegraphics[width=.8\textwidth]{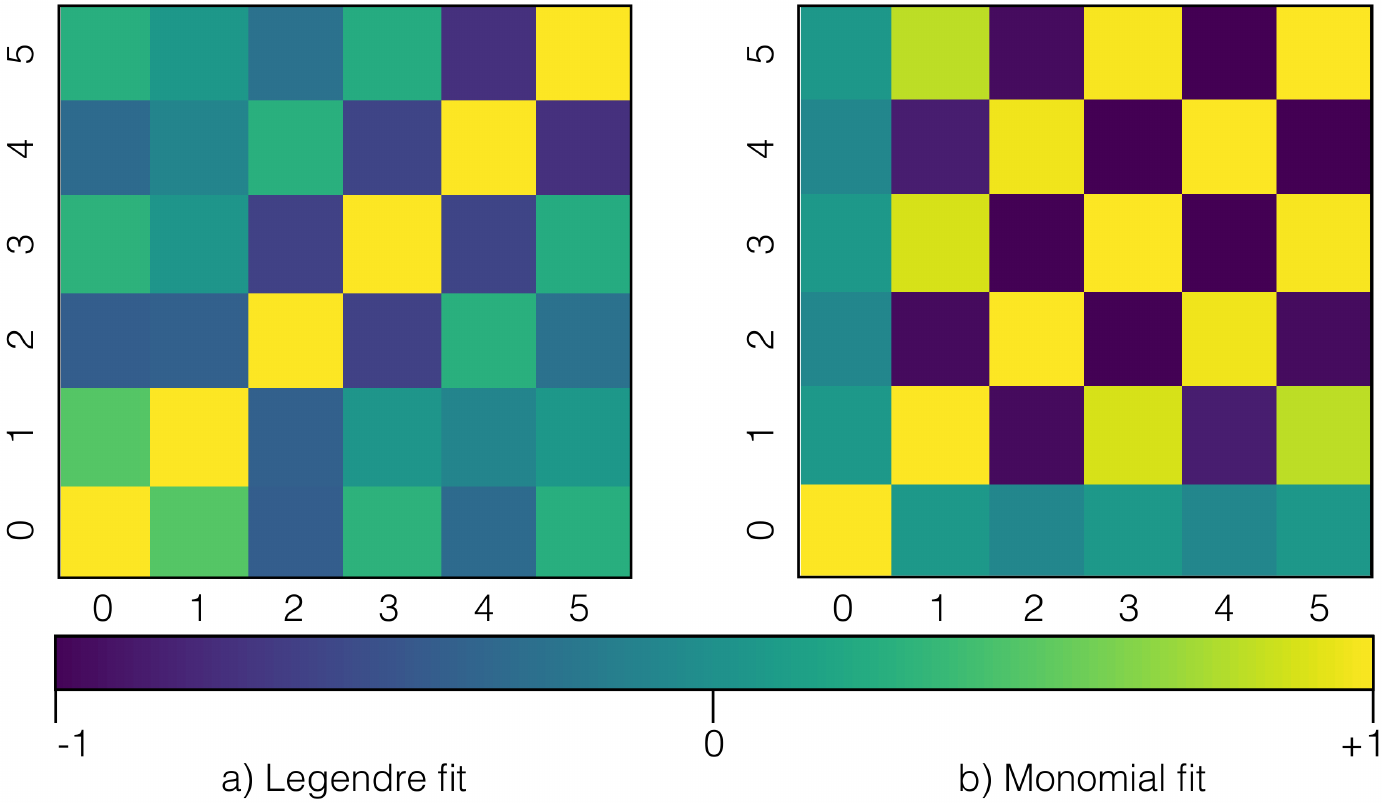}
\caption{We fitted the same Abell 370 data to $5^{th}$ degree (6 free parameters) using a) Legendre polynomials and b) monomials and computed the Pearson correlation matrices. The Legendre representation is strikingly less correlated than the monomials. As described in the text, the nearly diagonal correlation matrix of the Legendre polynomials is preferable to the checkerboard that the monomials produced.}\label{fig-corr}
\end{centering}
\end{figure*}

To compare the two representations, we fitted $\bnm{D}$ to $5^{th}$ degree using Legendre polynomials and the monomials. The nearly diagonal correlation matrix of the Legendre polynomials (Figure~\ref{fig-corr}a) is preferable to the checkerboard of the monomials (Figure~\ref{fig-corr}b).

There are at least two advantages to a diagonal covariance matrix. One advantage is that in order to do calculations, one needs to only know and deal with 6 components (the diagonals) instead of 21 components (including the off diagonals) for these observations.  This reduces the number of parameters that one needs to know a priori and saves computer space and computer time . The other, arguably more important, advantage is that inverting a large matrix with large off diagonal components leads to larger errors.  For monomials this leads to manifestly wrong answers with single precision arithmetic at $\text{deg}=6$ and even with double precision at $\text{deg}=11$.

\section{When Might One do Better?}\label{sec-special-cases}

The PCA shows that the Legendre polynomials are generally a good choice for modeling the response of sampled up-the-ramp pixels to astronomical scenes like Abell 370, the globular cluster 47 Tucanae (not discussed here), laboratory flats, and darks (when sufficient data are available). As discussed in $\S \ref{sec-why}$, one would expect the Legendre polynomials to emerge as a natural basis in many astronomical situations.

However, in some special cases, it might be possible to characterize and measure the actual eigenvectors. For example, this might be true for some transiting exoplanet observations. In these cases, the Legendre polynomials would not be a bad choice. However, one might expect to do even better by measuring and using the actual eigenvectors when it is practical to do so.

\section{Information Content by Legendre $\lambda$}\label{sec-info}

For Abell 370, Figure~\ref{fig-evals} shows that a $5^{th}$ degree Legendre fit extracts nearly all of the meaningful information. The only pixels that experience significant information loss are those that strongly saturate, are hit by cosmic rays, or are statistical outliers in some other way. For these rare events, techniques like PCA that rely on ensemble averages are not effective.

\begin{figure*}[t]
\begin{centering}
\includegraphics[width=\textwidth]{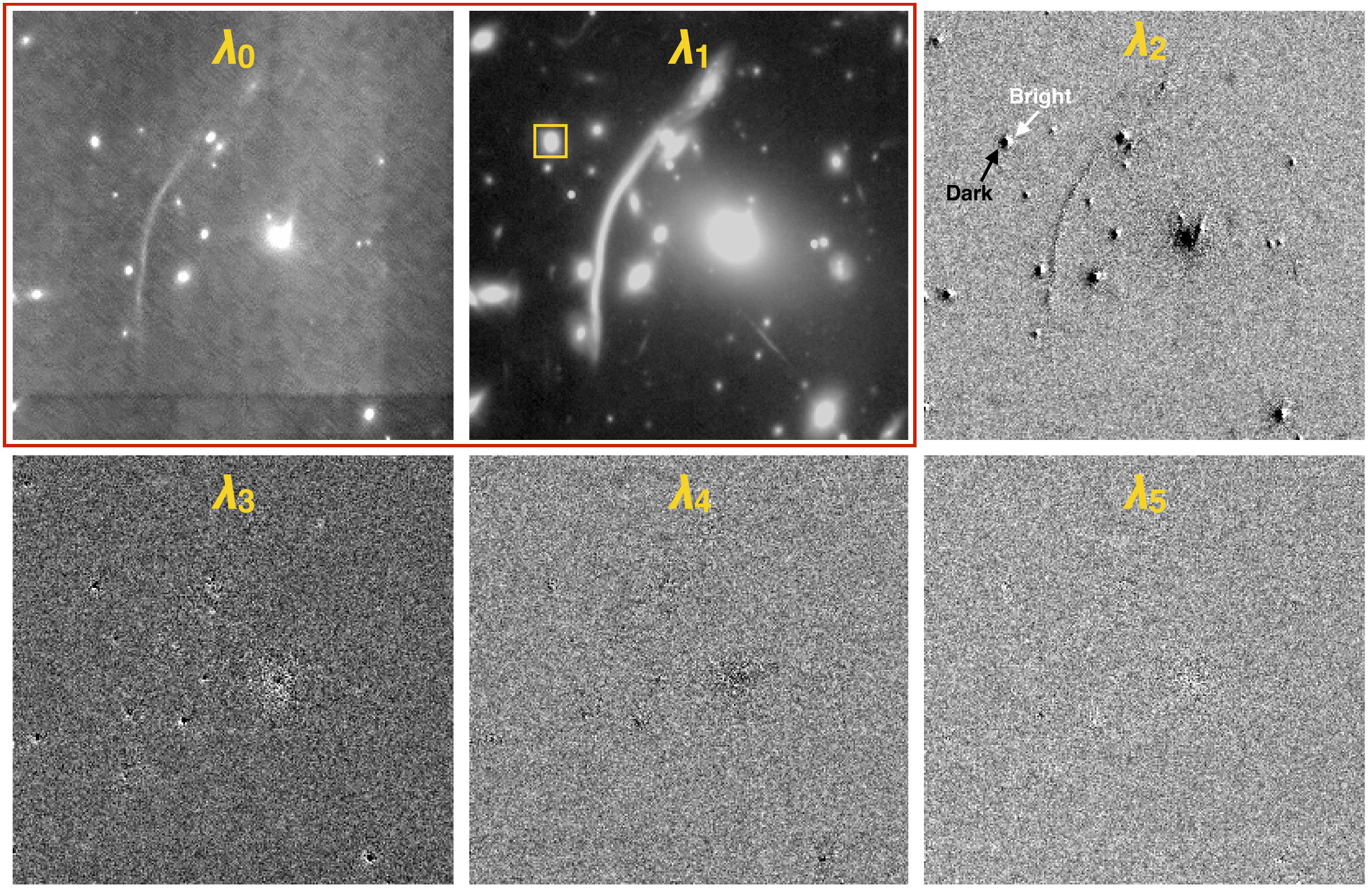}
\caption{This figure shows the six Legendre $\lambda$ coefficients for the ROI of Figure~\ref{fig-overview}. The red box highlightes the information that is used by the WFC3 calibration pipeline. After linearization, the pipeline fits a straight line ($\lambda_0$ and $\lambda_1$); thereby not utilizing the information that is contained in $\lambda_2-\lambda_5$ to constrain model parameters. Consistent with Figure~\ref{fig-evecs-leg}b, $\lambda_5$ and higher contain very little astronomical information. For comparison, we fitted the same data with a 2-parameter straight line. The yellow box is a photometer aperture. For this source, there is a 0.9\% difference in brightness between the $5^{th}$ degree Legendre fit and the straight line. The images show detector edges from stacking (especially $\lambda_0$). The bright-dark artifacts seen especially in $\lambda_2$ are more interesting. These may be caused by $\approx$1.5~milliarcsecond guiding errors (1\% of WFC3's pixel pitch) during each exposure. This is discussed more fully in Appendix~\ref{sec-bright-dark}}\label{fig-lambdas}
\end{centering}
\end{figure*}

Figure~\ref{fig-lambdas} provides a visual impression of how information content falls off with increasing $i$ in $\lambda_i$. This figure was made by computing the median of all 36 Legendre cubes after offsetting them to account for dithering. The offsets were computed by cross correlating the $\lambda_1$ (slope) coefficients. The integrated exposure time on the sky in each panel is about 14~hours.

$\lambda_0$ is the mean value of the up-the-ramp samples. It contains much of the instrument signature (bias pattern, hot pixels, bad pixels...), although there is also astrophysical information. Where there is a bright source, the mean value of the ramp is higher, so the sources are visible in $\lambda_0$.

$\lambda_1$ is proportional to the conventional slope image (the conversion factor is given in $\S \ref{sec-astro}$). As expected, it is strongly dominated by astronomical sources. As in conventional pipelines, $\lambda_1$ can be used directly for non-critical measurements.

$\lambda_2$ is the first term to capture curvature (see Figure~\ref{fig-evecs-leg}b). For nominal pixels that slowly lose response as they fill until entering saturation, it should always be negative when it exceeds the noise. $\lambda_2$ strongly reflects non-linearity. We anticipate that it will be extremely sensitive to anything that can impart curvature to the ramp. Some examples include intrinsic non-linearity for bright sources, non-linearity in the ROIC when/if bright sources perturb the ROIC's electrical state, the brighter-fatter effect, and pointing jitter and drifts.

The brighter-fatter effect can make a ``bull's-eye'' pattern around each bright star, with a dark central hole surrounded by a bright ring in the $\lambda_2$ image. The effect of drifting in/out from pixels would appear like the bright-dark artifacts seen in Figure~\ref{fig-lambdas}. Here $\lambda_2$ seems to have been sensitive to 1.5~milliarcsecond drifts (1\% of pixel pitch). For more information on the bright-dark pattern, please see Appendix~\ref{sec-bright-dark}.

The higher order $\lambda$s contain diminishing information. But, as can be seen in Figure~\ref{fig-lambdas}, information is still present in these Abell 370 data until at least $\lambda_5$. For disentangling time dependent instrument signatures, we believe that the time domain information contained in all six terms may be important.

\section{Information Compression}\label{sec-compress}

Legendre cubes are highly compressible. Looking at the Abell 370 data, consisting of the reset frame and 15 up-the-ramp samples, a complete data dump would be $16 \times 2$ = 32 bytes per pixel.  However, essentially all of the information can be retained in 6 Legendre coefficients. If these are stored as 4 byte floats this is 24 bytes per pixel, a modest 25\% compression. Compressing a typical astronomical data file with ``gzip --best'' results in 26.7 bytes/pixel, a 17\% compression.

However, floating point numbers are unnecessary and difficult to compress. Keeping 8 digits of precision on a number which has an uncertainty at the second digit is clearly excessive.  The ancient instruction of keeping a single digit into the uncertainty is good advice.  To update this to binary, one should keep 2 or 3 bits into the noise.  One way to do that is to convert the Legendre coefficients from floating point to integer (or calculate them that way to begin with), and then multiply them with a prearranged constant to make the 2 or 3 least significant bits be in the band of uncertainty. For astronomical data dominated by  poisson noise, the uncertainty is proportional to the square root of the signal. Taking the square root halves the number of bits required even before compression. If, as often the case, there is additional readout uncertainty, the proper treatment is to add a constant before taking the square root (see Ref.~\citenum{2005PASP..117...94O}, Fig.~1).

Here, the first two coefficients, $\lambda_0$ and  $\lambda_1$, are offset to make 
them positive and the square root results in data that can fit into 12 bits each 
per pixel, including three bits of noise. The $\lambda_2$ coefficient fits into 10 
bits and the remaining coefficients fit into 9 bits each including three bits of noise. 
Thus a full ramp can be packed into 60 bits or 7.5 bytes, a compression of 75\% or a 
factor of 4.  This file can furthermore be gzipped, resulting in a compression ratio of approximately $5-6\times$ relative to the starting data while retaining nearly all of the scientific information.

The final gzip compression of about 25\% is the best that can be expected since much of the ``data'' are the three bits of noise in each coefficient. If only two bits of noise are kept the data compresses to 4.6 bytes/pixel, for a total compression ratio of 7. Higher compression ratios can be expected for longer ramps. This is because even a 64 sample ramp on the \WFIRST H4RG-10 requires only $\approx$8 Legendre coefficients, and  these will require only a few extra bits for the coefficients. Table~\ref{Compression} provides a summary of the compression that can be achieved after the different steps described above.

\begin{table}[tbph]
\begin{center}
\caption{Compression Trades\label{Compression}}
\begin{tabular}{lcccc}
\hline
\multicolumn{5}{c}{Data in Various Forms}\\
\cline{2-4}
Data form & Item size & Number & byte per Pixel &Compression\\
\hline\hline
Original data&2 bytes&16 sample&32&-\\
Legendre Trans.&4 bytes&6 Coefs&24&25\%\\
Compress to int&10 bits avg&6&7.5&77\%\\
Zipped File& & & 5.6&83\%\\
Keep 2 bits of noise& & &4.6&87\%\\
\hline
\end{tabular}
\end{center}
\end{table}

\section{Astrophysics in Legendre Space}\label{sec-astro}

For many years, the standard practice in IR astronomy has been to linearize up-the-ramp sampled data and then fit straight lines to make a slope image.\cite{Vacca:2004fg,Hilbert:2014uy} Although the details of the implementations differ, the focus has been on collapsing 3D datacubes to 2D images. The 2D image, or more commonly averaged 2D images,  have almost always been the unit of data that is used for science analysis.

The PCA clearly shows that this approach does not optimally use the time domain information that is contained in the higher order terms. Using calibration files to linearize the data before fitting a straight line seeks to calibrate this information out, rather than use it to constrain model parameters. To the extent that the linearity correction is always imperfect, some time domain information is always left behind. To utilize all of the meaningful information (both astrophysical information and instrument signatures), one must retain more than just 2D slope images. In this regard, Legendre cubes are helpful because they approximate the eigenvectors.

As of today however, we are only just starting to learn how to work with Legendre cubes. As place holders for proper algorithms, we have found the following to be handy although imperfect tools for collapsing 3D Legendre cubes to familiar 2D images for debugging and comparison with legacy data products. Going forward, our aims include developing techniques for performing common analysis algorithms directly on Legendre cubes. We think that this will probably rely more on forward modeling than is common today. 

\subsection{Slope Estimators: Two Handy but Imperfect Tools}\label{sec-toolkit}

If a 2D image is desired, then $\lambda_1$ is proportional to the conventional slope image. If there are $n$ up-the-ramp samples and the frame readout time is $t_f$~seconds, then the slope in units of data numbers (DN) per second is,
\begin{equation}\label{eq-slope}
\text{slope}=\frac{2\lambda_1}{\left(n-1\right)}t_f^{-1}.
\end{equation}
If used as the only flux estimator, Equation~\ref{eq-slope} has the undesirable attribute that it does not optimally use the time domain information that is contained in the higher order terms.

Another handy way to get a 2D image that captures more of the information is to compute the integrated signal in DN using the Legendre fit,
\begin{equation}\label{eq-signal}
\text{integrated signal} = \sum_{i=0}^{\text{deg}}\lambda_i \left(P_i\left(+1\right)-P_i\left(-1\right)\right).
\end{equation}
Only the odd numbered terms contribute to the sum in Equation~\ref{eq-signal} because the even numbered Legendre polynomials are even functions. As such, some information is not optimally used. Compared to Equation~\ref{eq-slope}, Equation~\ref{eq-signal} has the advantage that it utilizes more of the available information and it is not much more difficult to compute. Equation~\ref{eq-signal} can straightforwardly be converted to the mean photocurrent during the exposure (units: DN per second),
\begin{equation}\label{eq-pc}
\text{mean photocurrent} = \sum_{i=0}^{\text{deg}}\frac{\lambda_i \left(P_i\left(+1\right)-P_i\left(-1\right)\right)}{(n-1)}t_f^{-1}.
\end{equation}

These image estimators are far from perfect. The first problem is that they collapse the 3D Legendre cube to a 2D image before performing any scientific analysis. In doing this, resolution on the time domain information is lost. However, the time domain may hold the keys to disentangling source flux from persistence, burn in, and other detector artifacts mediated by charge trapping and release.

A second well known problem is that the detectors are inherently non-linear. Non-linearity must be accounted for when mapping instrumental brightness in DN to astronomical source brightness. As has already been discussed, today's pipelines use calibration data to correct for non-linearity before slope fitting (or vice versa). These operations generally do not commute: that is to say that if slope fitting is done before linearization one gets a slightly different result. Our eventual aim is to account for non-linearity simultaneously with measuring source brightness as part of fitting (or forward modeling) a multi-parameter model to the pixel's response as recorded in the Legendre cube. However, this is work for the future.

\subsection{Plans to Study Linearity}\label{sec-linearity}

Given the lack of a satisfactory network of faint standard stars spanning a range of brightness and color, we understand that we must learn how to model non-linearity in Legendre cubes. To address this, we are pursuing two parallel tracks.

One track is purely experimental. We are upgrading a laboratory test setup at Goddard so that we can apply light of known wavelength and relative brightness over a wide, $10^{4-5}\times$ dynamic range. The aim is to use this to study the linearity properties of Legendre fitted pixels. If an opportunity were to come up to fly a similarly calibrated detector in a very well understood telescope, that would be invaluable in establishing a network of faint astronomical calibrators with known relative brightnesses.

The second track combines theory and experiment. By experimentally studying the properties of individual pixels and building physical models of them,  we hope that it may be possible to apply valid linearity corrections using less calibration data. However, work is just beginning on the theoretical modeling. The test setup described above is still under construction. We look forward to saying more about these topics as things mature.

\section{Future Work}\label{sec:future}

Our hypothesis going forward is that for \WFIRST's most demanding surveys, it may be beneficial to replace linearization followed by line fitting with fitting (or forward modeling) a multi-parameter model to Legendre cubes. Linearization would notionally be part of this process, as would measuring source brightness, and accounting for artifacts such as persistence. 

For space, we must also understand how to recognize cosmic rays in Legendre cubes. Work on this is just beginning. However, it may be that the Legendre cubes themselves contain sufficient information to differentiate cosmic ray hits from normal pixel response. In other words, cosmic ray hits may leave a distinct signature in Legendre space that differs from normal pixel response. This is something that we can begin to explore with existing laboratory data from \JWST and \WFIRST and flight data from WFC3.

\section{Summary}\label{sec-summary}

In this paper, we reported the results of PCA of laboratory flats and real astronomical data from a variety of IR instruments. The detectors included WFC3's Teledyne H1R, \JWST's H2RGs, and \WFIRST H4RG-10s. The IR array controllers included WFC3's flight electronics, \JWST SIDECAR ASICs, and Gen-III Leach controllers from Astronomical Research Cameras, Inc., for \WFIRST. In all cases, the Legendre polynomials emerged as a better basis (in the linear algebra sense) for modeling up-the-ramp sampled pixels than the monomial basis that is used for polynomial fitting and linearization in today's calibration pipelines.\footnote{Some colleagues have suggested that the Fourier basis might be a good choice (after projecting out the offset and slope terms). While preparing this manuscript, we tried doing this. Compared to the Legendre polynomials, the Fourier basis is also a poor choice.} The Legendre polynomials are an orthogonal basis whereas the monomials are not. Moreover, the PCA shows that the Legendre polynomials are often a reasonable proxy for the eigenvectors that diagonalize the covariance matrix whereas the monomials are not. As such, the Legendre polynomials provide a less correlated representation of the data than the conventional monomials.

Building on the PCA, in $\S \ref{sec-compress}$ we described how the information content in astronomical scenes falls off with fit degree. For the \HST WFC3 Frontier Field observations of Abell~370, we showed that nearly all of the meaningful information could be compressed from 15 up-the-ramp samples down to 6 Legendre coefficients per pixel. Moreover, Legendre coefficients are compressible. In $\S \ref{sec-compress}$, we describe concepts for compressing data like these down to 14\% of the raw data volume while retaining nearly all of the meaningful information. 

Looking to the future, we believe that a change of basis from the monomials to the Legendre polynomials should benefit not only compression, but also IR array calibration. The PCA shows that up-the-ramp sampled data contain more information than is contained in the conventional offset and slope images. The additional information exists in the time domain, and we believe that it may be important for disentangling time dependent artifacts like inherent non-linearity, brighter fatter effect, persistence, and pointing drifts, \etc Linearizing the up-the-ramp samples before fitting slopes, as is done today, does not use the time domain information to constrain model parameters. Rather, it seeks to remove it. Our hypothesis going forward is that it may be possible to do better by directly fitting, or forward modeling, multi-parameter models to Legendre cubes.

\appendix

\section{WFC3 H1R Detector, Up-the-ramp Sampling, and Reference Correction}\label{sec-detector-info}

\subsection{WFC3 H1R Detector}\label{sec-format}

\begin{figure*}[t]
\begin{centering}
\includegraphics[width=0.8\textwidth]{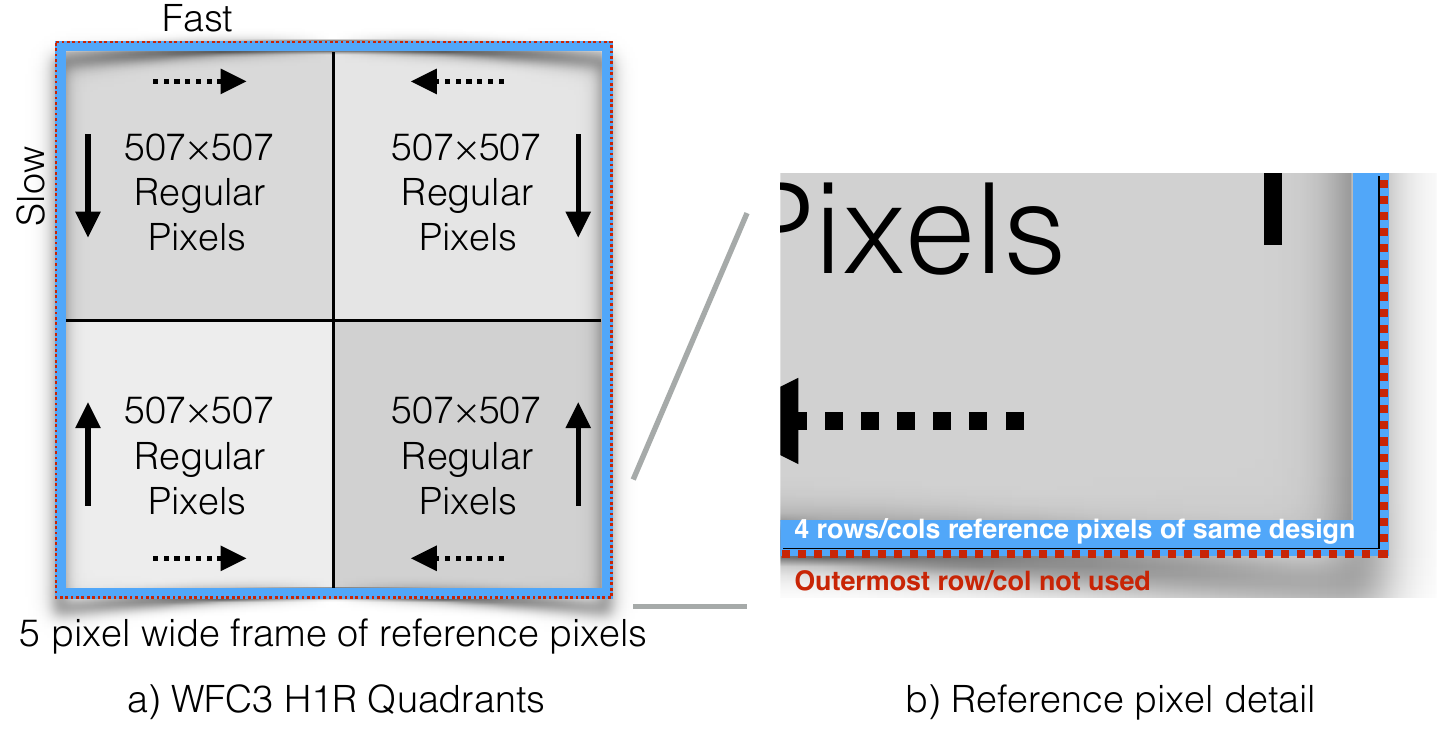}
\caption{WFC3's H1R detector is read out in quadrants. The full array has 1024$\times$1024 pixels, of which 1014$\times$1014 are regular photosensitive pixels. These are surrounded on all sides by a 5 pixel wide border of reference pixels. The outermost row/column of reference pixels was not used because it contains several different reference pixel designs. The inner 4 rows/columns were used because these reference pixels are all of the same design.}\label{fig-h1r}
\end{centering}
\end{figure*}

WFC3 IR uses a Teledyne H1R near-IR detector array (Figure~\ref{fig-h1r}a). The H1R is a 1024$\times$1024~pixel HgCdTe detector array. Pixels are read out in four 512$\times$512~pixel ``quadrants''. The HgCdTe is tuned to have a cutoff wavelength of about $1.7~\mu$m. The ``R'' in its name denotes reference pixels.

The H1R was the first astronomical near-IR detector to provide engineered reference pixels embedded in the video outputs. In the H1R, the reference pixels appear in a 5 pixel wide band framing the 1014$\times$1014 photosensitive pixel area (Figure~\ref{fig-h1r}b). Although the reference pixels are designed to electronically mimic a regular pixel, they do not respond to light. During calibration, they are used to remove electronic drifts as described in $\S \ref{sec-refcor}$.

\subsection{IR Array Readout and Up-the-ramp Sampling}\label{sec-utr}

Unlike in a CCD, it is possible to non-destructively sample the pixels in H1R and HxRG (and similar IR array) detectors many times without resetting them. At first, ``multiple non-destructive reads'' were used to enable ``Fowler sampling'', which quickly samples the detector several times at the beginning and end of each integration to ``average down'' the read noise. Fowler and Gatley (1991)\cite{1991SPIE.1541..127F} still provides a good introduction to Fowler sampling and other uses of multiple non-destructive reads.

\begin{figure}[t]\begin{centering}
\includegraphics[width=.5\textwidth]{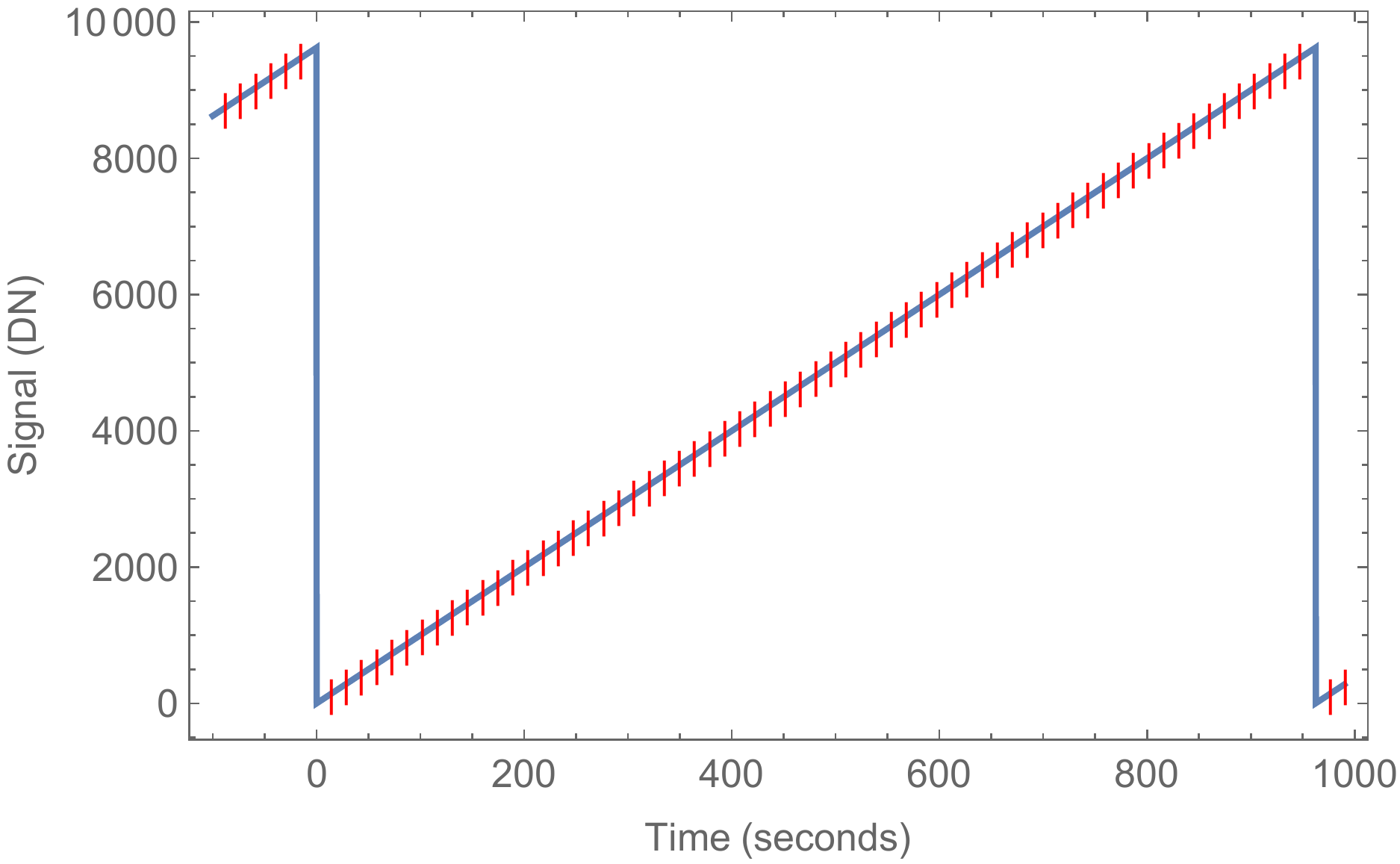}
\caption{This figure shows up-the-ramp sampling for one pixel as it is implemented in the \JWST NIRSpec. There are 65 non-destructive samples per $\textrm{EXPTIME}=933~s$ exposure. To maintain constant power dissipation, the detector is clocked and pixels are digitized at a constant cadence. For the NIRSpec mode that is shown here, that cadence is about 14.6 seconds per frame. Charge integration begins upon the completion of reset for the previous exposure ($t=0~s$). Each vertical red line indicates a sample. Note that some signal has integrated by the time the pixel is digitized for the first time. The intent of this figure is to show the timing from the perspective of one NIRSpec pixel. \JWST observers should see the NIRSpec information pages (https:\/\/jwst-docs.stsci.edu\/display\/JTI\/Near+Infrared+Spectrograph\%2C+NIRSpec) for technical information about actual \JWST data products. Although the details differ slightly for WFC3 and \WFIRST, the underlying concept of acquiring multiple non-destructive samples is the same.}\label{fig-utr-sampling}
\end{centering}\end{figure}

Building on these ideas (and briefly mentioned in Fowler and Gatley), it is possible to non-destructively sample a detector uniformly throughout an exposure. Today, this is known as up-the-ramp sampling. Up-the-ramp sampling is widely used in space because cosmic ray disturbance is easily recognized and can sometimes be corrected for. Figure ~\ref{fig-utr-sampling} shows an example of up-the-ramp sampling as it is implemented in \JWST NIRSpec.

In all IR array readout modes that use multiple non-destructive samples, there are correlations between the samples within an exposure that affect the uncertainty in the measured flux. Garnett and Forrest (1993)\cite{Garnett:1993vi} provides a good introduction to correlated noise trades as a function of readout mode. Vacca, Cushing, and Rayner (2004)\cite{Vacca:2004fg} provides an updated discussion that includes the effects of detector non-linearity. Rauscher (2010; Equation~1)\cite{2010PASP..122.1254.} and Rauscher \etal (2007)\cite{Rauscher:2007gc} provides an update to Vacca \etal's noise model for the special case of \JWST readout.

\subsection{Reference Correction}\label{sec-refcor}

WFC3's H1R detector is read out in four ``quadrants'' (Figure~\ref{fig-h1r}a). The quadrants can (and do) have DC offsets with respect to each other. The reference pixels are used to take these out in post-processing. The outermost 5 pixels on all sides are reference pixels. Of these, the outmost rows/columns contain reference pixel design variations. The inner four rows/columns of reference pixels all use the same design and these are the ones that we used for this study.

We used a very simple reference correction scheme for this study. The reference correction was applied to each frame individually as the first processing step. We treated each quadrant separately, and computed the median of all reference pixels in each quadrant excluding the outermost rows/columns. This median was then subtracted from every pixel in that quadrant. The end result was a DC reference correction that was applied frame-by-frame and quadrant by quadrant. After making the reference pixel correction, we cropped all reference pixels from each frame.

More sophisticated reference correction is possible given the right hardware, clocking patterns, and calibration software. For example, some instruments treat the reference rows and columns separately. With careful tuning, it is sometimes possible to use the reference columns to remove noise within the outputs (not just DC). For \JWST NIRspec, we developed Improved Reference Sampling and Subtraction (\IRSSquare; pronounced ``IRS-square'').\cite{2017PASP..129j5003R} \IRSSquare uses a specialized clocking pattern to interleave many more reference pixels into the data than is otherwise possible, resulting in cosmetically cleaner images with less correlated noise.

In any case, the conclusions of this paper appear to be robust against changes in how the reference pixels are used. In addition to the very simple DC correction described above, we also did PCA on \IRSSquare sampled flats (Appendix~\ref{PCA-of-flats}). In both cases, PCA showed that the Legendre polynomials were a good approximation to the eigensystem.

\section{PCA of \JWST and \WFIRST Flats}\label{PCA-of-flats}

\begin{figure*}[t]
\begin{centering}
\includegraphics[width=.5\textwidth]{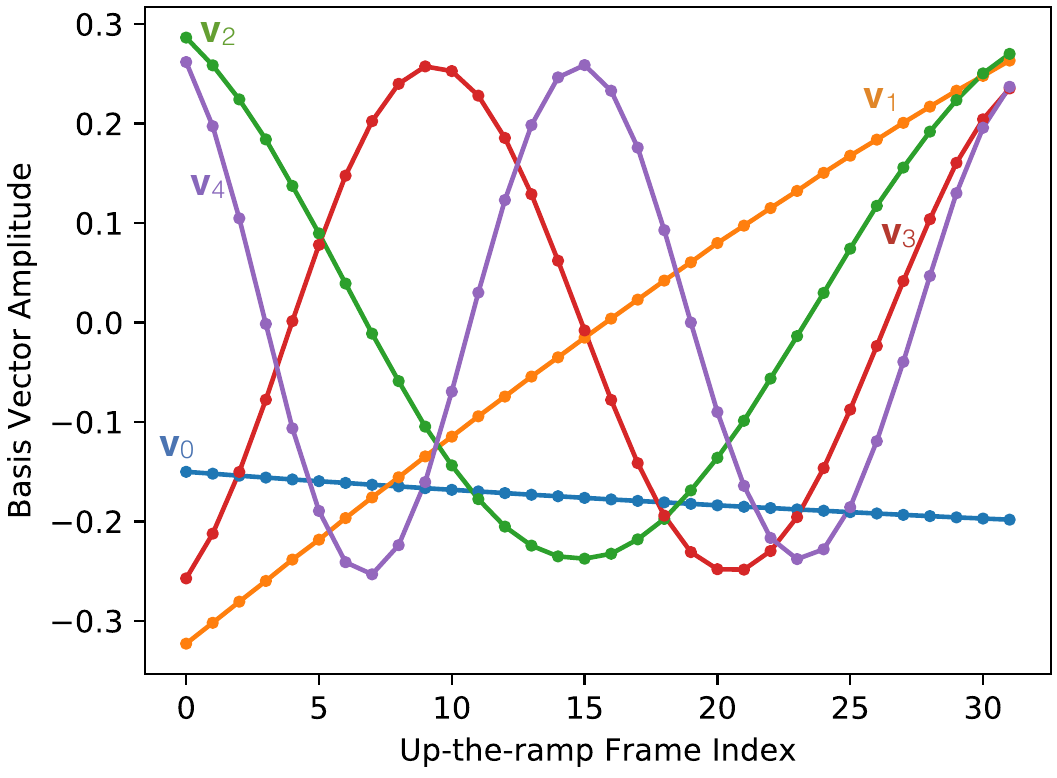}
\caption{The eigenvectors, $\bnm{v}$, of a \JWST NIRSpec channel 491 flatfield exposure to nearly full well are very similar to the Legendre polynomials. \WFIRST flats show similar behavior.}\label{fig-jwst-pca}
\end{centering}
\end{figure*}

The Legendre polynomials are a good choice of basis for both astronomical scenes and flats. Figure~\ref{fig-jwst-pca} shows the eigenvectors for a \JWST NIRSpec flatfield image exposed to nearly full well ($6\times 10^4~e^-$). The procedure was as described in $\S \ref{sec-pca}$. By inspection, the eigenvectors are similar to the Legendre polynomials (Figure~\ref{fig-evecs-leg}b). We have repeated this analysis for flats taken with \WFIRST H4RG-10s and gotten similar results.

\section{Read Noise Distribution at High $\lambda_i$}\label{sec-gaussian-noise}

If all of the information were captured by $\lambda_0-\lambda_5$, then the distribution in $\lambda_6$ and higher would be Gauss-normal noise. To test this hypothesis, we used three standard statistical tests: (1) the Kolmogorov-Smirnov (KS) test, (2) Q-Q plots, and (3) the Anderson-Darling test. The KS test shows that by $\lambda_6$, the noise is statistically indistinguishable from gauss-Normal. The Q-Q plots also showed this to be true, except for statistical outliers. The Anderson-Darling test weights statistical outliers heavily, so it shows that although the distribution is trending toward Gauss-normal, there are nevertheless too many statistical outliers to be well represented by an underlying Gauss-normal distribution. However, if we clip statistical outliers, then the Anderson-Darlington test, like the Q-Q plots and KS test becomes consistent with an underlying Gauss-normal distribution. In a statistical sense, for these \HST data, essentially all of the meaningful information is captured by $\lambda_0-\lambda_5$ once a very small sample of outliers has been eliminated.

\subsection{KS Tests}\label{sec-ks}

As the information content (\ie variance) decreases in the 
higher order $\lambda_i$ (Figure~\ref{fig-evals}), we also expect that the distributions of 
$\lambda_i$ values should become become increasingly well represented
by a Gauss-normal distribution. Thus a one-sided KS test (see Ref.~\citenum{Press:1992:NRC:148286}, $\S 14.3$)
was applied to compare the $\lambda_i$ values to a Gaussian function 
for each of 111 ramps (36 at F160W, the rest using other filters).
The KS test calculates the maximum deviation ($D$) between the 
cumulative distribution functions for the data and for the Gaussian model. 
Then, one calculates the probability for finding a value of $\geq D$ under
the assumption that the data are drawn from a Gauss-normal distribution. Low 
probabilities indicate that $D$ is improbably large for data with 
a truly Gauss-normal 
distribution. Over many repeated tests of Gauss-normal data 
(with known mean and $\sigma$ values), the probabilities
should fall in a uniform distribution over the range [0,1].
Because of the detector readout pattern and the reference correction 
(Appendix~\ref{sec-detector-info}), the 4 detector quadrants were tested independently.

\begin{figure}[h] 
   \centering
   \includegraphics[width=3in]{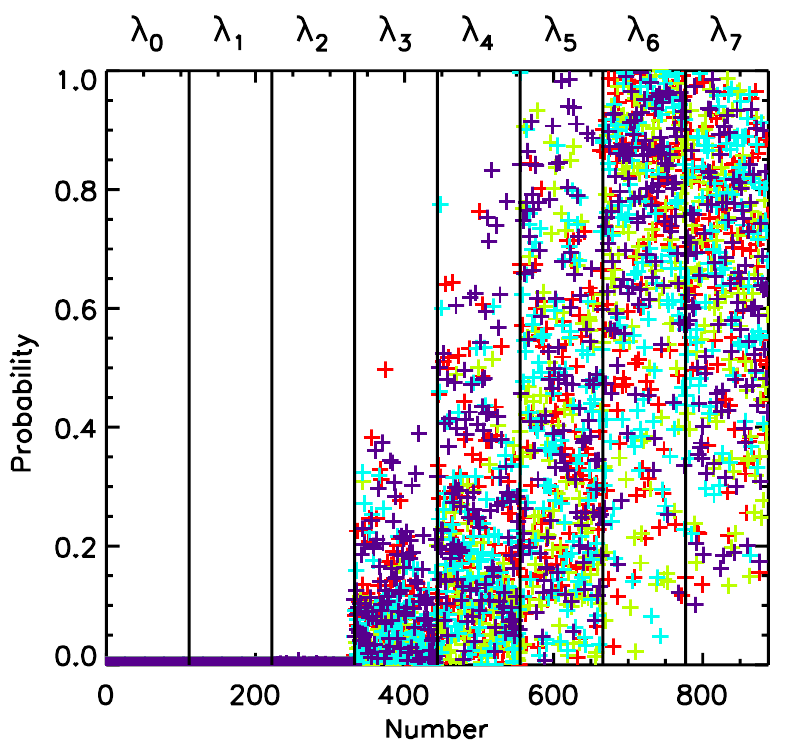} 
   \caption{One-sided KS probabilities for testing the Gauss-normal 
   distribution of $\lambda_i$. The horizontal axis simply disperses the 
   different $\lambda_i$ and 111 ramps. The lowest $\lambda_i$ carry information 
   and are strongly non-Gaussian. $\lambda_6$ and $\lambda_7$ are 
   consistent with Gauss-normal distributions, accounting for the bias 
   produced by fitting the mean and $\sigma$ of each distribution rather 
   than comparing to a priori fixed values. Symbol colors distinguish the 
   independently-tested 4 quadrants for each $\lambda_i$.
   \label{fig:ks}}
\end{figure}

Figure \ref{fig:ks} shows the KS probabilities for each of 111 $\lambda_i$
values, in cases where $n=7$ Legendre components were fit. There is zero chance 
that $\lambda_0$ through $\lambda_2$ are Gauss-normal 
distributions. However, the probability of Gauss-normal distributions 
increases from $\lambda_3$ through $\lambda_5$. For $\lambda_6$ and 
$\lambda_7$ the distributions are indistinguishable from Gauss-normal. The 
apparent bias towards high probabilities is a result of fitting the mean and 
$\sigma$ for each of the tests, as these parameters are not known a priori.
In summary, for these Abell 370 data, the KS test 
suggests little or no information is contained beyond $\lambda_5$.

\subsection{Q-Q Plots}\label{sec-qq}
A quantile-quantile (Q-Q) plot is a graphical diagnostic to test whether two samples come from the same distribution. They plot one sample's quantiles against another sample's quantiles. If the samples come from the same distribution, the points lie along a 1:1 line. Here, what is meant by a quantile is simply a data point below which a certain proportion of the data falls. To test whether the noise distributions in the higher $\lambda_i$ terms are Gauss-normal, we can compare the sample quantiles of the $\lambda_i$ data by quadrant to the theoretical quantiles of a Gaussian normal distribution.

\begin{figure*}[t]
\begin{centering}
\includegraphics[width=\textwidth]{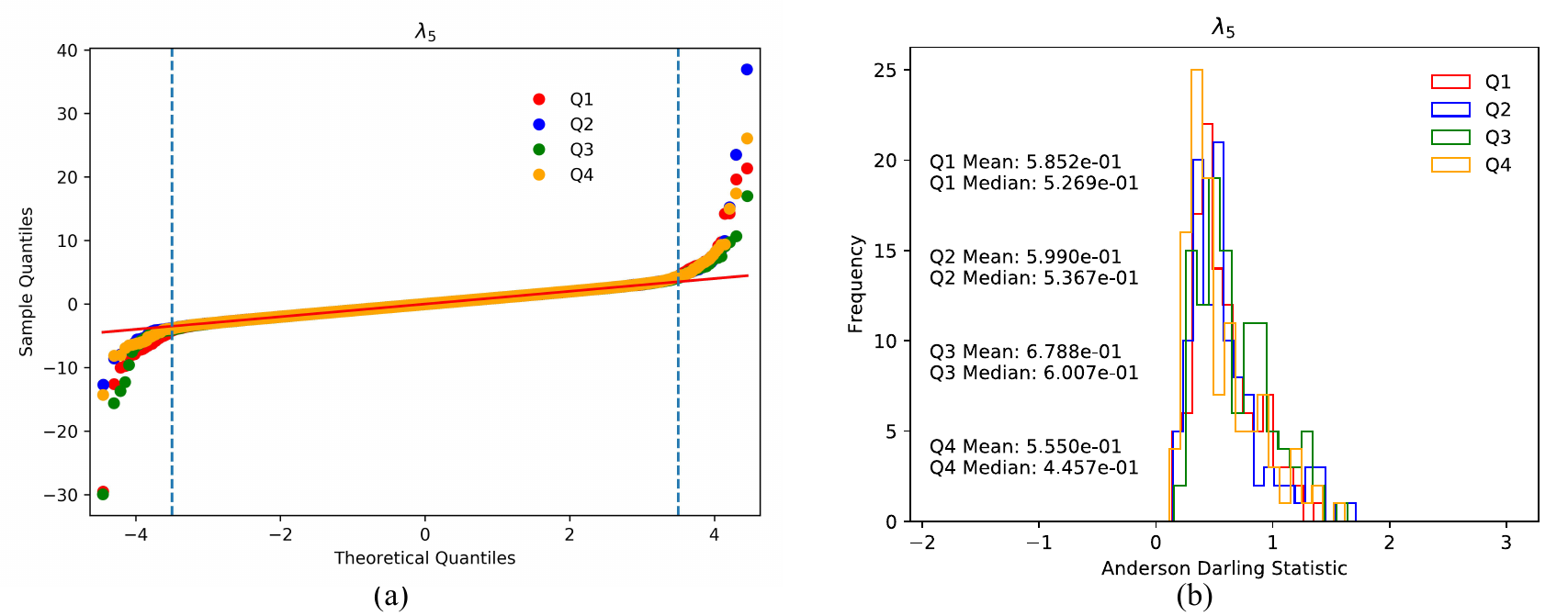}
\caption{(a) Q-Q plot comparing $\lambda_5$ data with a Gauss-normal distribution with each quadrant represented by a different color of points. The $\lambda_5$ data have had the mean value subtracted and been divided by the sample standard deviation. A 1:1 line is shown in red. Higher order terms tend to approach the 1:1 line implying increasing normality, but heavy tails do exist at the extrema. Dashed lines mark 3.5 standard deviations from the mean where it seems the data starts to deviate from normality in the higher order terms. (b) Histogram of the Anderson-Darling statistic for $\lambda_5$ excluding the $>$3.5 standard deviation wings (about 1\% of pixels) with different quadrants represented by different colors. The mean and median of the statistic for each quadrant are printed in the plot. All means are less than 1.092 suggesting the clipped data are consistent with a Gauss-normal distribution.}\label{fig-qqadplot}
\end{centering}
\end{figure*}

Figure~\ref{fig-qqadplot}a shows the resulting Q-Q plots for $\lambda_5$ with each quadrant represented by a different color of points. The $\lambda_5$ data has had the mean value subtracted and been divided by its sample standard deviation. A 1:1 line is shown in red. In this plot, it is clear that data in this higher $\lambda_i$ term is mostly aligned with the 1:1 line suggesting these data approach Gaussian normality at higher order. However, the data deviate from normality at the extremes. These deviations are the results of heavy tails in the distribution of the $\lambda_i$ values which could be due to cosmic rays or other uncorrected effects. Examining the higher order $\lambda_5-\lambda_7$ Q-Q plots, we see the significant deviations from normality occur at about 3.5 standard deviations from the mean. Dashed lines show this limit, and if we remove these data we see that the data at higher order become more consistent with a Gauss-normal distribution as measured quantitatively with the Anderson-Darling tests described in \ref{sec-ad}.

\subsection{Anderson-Darling Test}\label{sec-ad}
The Anderson-Darling test tests the null hypothesis that the given sample data is drawn from a specific distribution.\cite{Anderson:1952fw} The Anderson-Darling test returns the Anderson-Darling statistic that can be compared to tabulated critical values (given a chosen significance level) to assess the validity of the null hypothesis\cite{Stephens:1974}. The Anderson-Darling test, like the KS test, compares data distributions, however, the Anderson-Darling statistic is calculated applying weights that make the Anderson-Darling test sensitive to differences in the beginning and ends of distributions \ie the tails. We select a significance level of 1\% corresponding to a 1 in 100 chance of rejecting Gaussian normality when the data are actually Gaussian normal. The corresponding critical value for the Anderson-Darling statistic when comparing to the normal distribution for a large sample size is 1.092.

We calculate the Anderson-Darling statistic for the $\lambda_i$ data for each quadrant across 111 images for data fit to 6 Legendre polynomials. Without removing the outliers in the tails of the data distribution marked by the dashed lines in Fig.~\ref{fig-qqadplot}a, the Anderson-Darling statistics for all images are greater than 1.092 suggesting a non-Gaussian normal distribution. However, if we exclude the data in the wings of the distribution, on average the Anderson-Darling statistics become less than one suggesting the data is consistent with a Gauss-normal distribution. Figure~\ref{fig-qqadplot}b shows the histogram of the Anderson-Darling statistic for $\lambda_5$ excluding the $>$3.5 standard deviation wings with different quadrants represented by different colors. The mean and median of the statistic for each quadrant are printed in the plot. Accounting for the outliers in the tails of the data distribution, the data suggest that at higher order the coefficients in the Legendre fit become Gauss-normal.

\section{Bright-Dark Artifacts in $\lambda_2$}\label{sec-bright-dark}

As $P_2(x)$ is the first non-linear Legendre polynomial, we expected that $\lambda_2$ would be strongly imprinted with the non-linear detector response found at very bright sources. As such, we expected $\lambda_2$ to be negative and strongly biased towards the brightest sources. However, Figure~\ref{fig-lambdas} reveals a bright-dark artifact such that $\lambda_2$ has both negative and positive values on opposite sides of bright sources.

We hypothesize that these artifacts may be caused by a drift in the WFC3 pointing during the course of the observations. Where a source is drifting into a pixel, the accumulating up-the-ramp signal will steepen with time, requiring a positive value of $\lambda_2$. Conversely, as a source drifts away from a pixel, the ramp will flatten (as it would from a standard non-linear response), and $\lambda_2$ will be negative. For small shifts, the induced values of $\lambda_2$ should be the product of the gradient of the brightness and the amount of the drift.  

To test this, we approximated the brightness gradient by differencing images of the median $\lambda_1$ shifted by $(-\delta x,-\delta y)$ and by $(+\delta x, +\delta y)$. To match the apparent direction of the shift in the $\lambda_2$ image, the actual gradient image used here is the average of gradient images made with $(\delta x, \delta y) = (1,1)$ and $(\delta x, \delta y) = (1,0)$. This approximates a gradient image of $\lambda_1$ in a direction that is $30\arcdeg$ from the $x$ axis ($d\lambda_1/dz$ where $\hat{z} = (2\hat{x}+\hat{y})/\sqrt{5}$). Next we performed a linear correlation between $\lambda_2$ and $d\lambda_1/dz$ for 6728 pixels with significant but not saturated signals. The slope of this correlation provides a measure of the fractional pixel shift. This shift translates to 1.54~mas given the 130~mas pixel size of for the WFC3 IR detector. The subtraction of the scaled $\lambda_1$ gradient leaves a $\lambda_2$ residual where the imprints of astronomical sources are almost entirely negative, and are tightly (but not linearly) correlated with $\lambda_1$, as would be expected for standard non-linear response.

The amount of drift during each of the 36 ramps was also estimated from examination of the RA and DEC data in the associated jitter files. The systematic drifts during each ramp are quite small (comparable to the standard deviation), and averaged across 36 ramps also had a value of 1.54~mas. The directions of the drifts were similar for all ramps, which was likely essential to their systematic imprint in the median $\lambda_2$. However,  the direction of the drift implied by the most straightforward interpretation of the jitter files is nearly $90\arcdeg$ different from the apparent drift in the data. When asked, former members of the WFC3 IR Development Team at Goddard told us that the recorded jitter was not necessarily for the specific line of sight to our field, and that known (or suspected) flexures and offsets in the observatory could potentially explain the discrepancy. The status is therefore that the drift amplitude implied by the bright-dark artifacts is consistent with the amplitude recorded in the jitter files. However, we were not able to reconcile the apparent drift directions with the jitter files. 

If one accepts that the bright-dark artifacts are consistent with pointing drifts, then a small drift as implied by $\lambda_2$ would necessarily distort the shapes of sources measured in that data. Use of the information in $\lambda_2$ to try to deconvolve the image to get the true source shapes is likely to be unstable. A better use of this information is to incorporate the drift into models that are compared to the data, so that both the data and model are drifted and comparisons between the two will be unbiased by the effect.

\acknowledgments 
This work was supported by NASA as part of the James Webb Space Telescope (\JWST), Wide Field Infrared Survey Telescope (\WFIRST),  and Hubble Space Telescope (\HST) Projects. We are   grateful for generous support over many years from the Goddard Internal Research and Development (IRAD)  program. We wish to thank the Goddard Detector Characterization Laboratory (DCL) for their superb work. Our first attempts at PCA used \WFIRST H4RG-10 test data. \WFIRST scientist Bob Hill, \WFIRST manager Dave Content, and DCL detector engineers Roger Foltz, Yiting Wen, Laddawan Miko, and Augustyn Waczynski (and  others) made these outstanding data  possible. We confirmed the \WFIRST findings using DCL data from the \JWST NIRSpec Detector Subsystem ground calibration campaign. In addition to those already mentioned, \JWST Scientist Matt Greenhouse, DCL engineer Brent Mott, software engineer Donna Wilson, and SIDECAR ASIC designer Markus Loose contributed to making the NIRSpec test campaign the success that it was. When it was time to move on to real astronomical data, the STScI archive was a pleasure to use. The Abell 370 data were obtained from the Mikulski Archive for Space Telescopes (MAST). STScI is operated by the Association of Universities for Research in Astronomy, Inc., under NASA contract NAS5-26555. We wish to thank Stephan Birkmann and Pierre Ferruit of the \JWST NIRSpec team for reading the entire manuscript prior to submission and providing valuable comments. We wish to thank the anonymous reviewers for carefully reading the manuscript and providing helpful comments that have made the paper more accessible to a wider audience.

\bibliography{ms}   

\begin{thebibliography}{10}

\bibitem{2015arXiv150303757S}
D.~{Spergel}, N.~{Gehrels}, C.~{Baltay}, D.~{Bennett}, J.~{Breckinridge},
  M.~{Donahue}, A.~{Dressler}, B.~S. {Gaudi}, T.~{Greene}, O.~{Guyon},
  C.~{Hirata}, J.~{Kalirai}, N.~J. {Kasdin}, B.~{Macintosh}, W.~{Moos},
  S.~{Perlmutter}, M.~{Postman}, B.~{Rauscher}, J.~{Rhodes}, Y.~{Wang},
  D.~{Weinberg}, D.~{Benford}, M.~{Hudson}, W.~S. {Jeong}, Y.~{Mellier},
  W.~{Traub}, T.~{Yamada}, P.~{Capak}, J.~{Colbert}, D.~{Masters}, M.~{Penny},
  D.~{Savransky}, D.~{Stern}, N.~{Zimmerman}, R.~{Barry}, L.~{Bartusek},
  K.~{Carpenter}, E.~{Cheng}, D.~{Content}, F.~{Dekens}, R.~{Demers},
  K.~{Grady}, C.~{Jackson}, G.~{Kuan}, J.~{Kruk}, M.~{Melton}, B.~{Nemati},
  B.~{Parvin}, I.~{Poberezhskiy}, C.~{Peddie}, J.~{Ruffa}, J.~K. {Wallace},
  A.~{Whipple}, E.~{Wollack}, and F.~{Zhao}, ``{Wide-Field InfrarRed Survey
  Telescope-Astrophysics Focused Telescope Assets WFIRST-AFTA 2015 Report},''
  {\em ArXiv e-prints} , arXiv:1503.03757  (2015).

\bibitem{Shapiro:2013gx}
C.~Shapiro, B.~T.~P. Rowe, T.~Goodsall, C.~Hirata, J.~Fucik, J.~Rhodes,
  S.~Seshadri, and R.~Smith, ``{Weak Gravitational Lensing Systematics from
  Image Combination},'' {\em PASP} {\bf 125}, 1496--1513  (2013).

\bibitem{Plazas:2018bw}
A.~A. Plazas, C.~Shapiro, R.~Smith, E.~Huff, and J.~Rhodes, ``{Laboratory
  Measurement of the Brighter-fatter Effect in an H2RG Infrared Detector},''
  {\em PASP} {\bf 130}, 065004--13  (2018).

\bibitem{Moore:2006kz}
A.~C. Moore, Z.~Ninkov, and W.~J. Forrest, ``{Quantum efficiency overestimation
  and deterministic cross talk resulting from interpixel capacitance},'' {\em
  Optical Engineering} {\bf 45}(7), 076402  (2006).

\bibitem{MacKenty:2008kj}
J.~W. MacKenty, R.~A. Kimble, R.~W. O'Connell, and J.~A. Townsend, ``{Wide
  Field Camera 3: science capabilities and plans for flight operation},'' in
  {\em Proc SPIE},  J.~M. Oschmann~Jr, M.~W.~M. de~Graauw, and H.~A. MacEwen,
  Eds., 70101F--9, Proc SPIE  (2008).

\bibitem{MacKenty:2010jd}
J.~W. MacKenty, R.~A. Kimble, R.~W. O'Connell, and J.~A. Townsend, ``{On-orbit
  performance of HST Wide Field Camera 3},'' in {\em Proc SPIE},  J.~M.
  Oschmann~Jr, M.~C. Clampin, and H.~A. MacEwen, Eds., 77310Z--12, Proc SPIE
  (2010).

\bibitem{MacKenty:2012ky}
J.~W. MacKenty, ``{Performance and calibration of the HST Wide Field Camera
  3},'' in {\em Proc SPIE},  M.~C. Clampin, G.~G. Fazio, H.~A. MacEwen, and
  J.~M. Oschmann, Eds., 84421V--10, Proc SPIE  (2012).

\bibitem{Dressel:2018}
L.~Dressel, {\em {The WFC3 Instrument Handbook, Version 10.0}}.
\newblock Baltimore: STScI, 10.0~ed.  (2018).

\bibitem{Gardner:2006di}
J.~P. Gardner, J.~C. Mather, M.~Clampin, R.~Doyon, M.~A. Greenhouse, H.~B.
  Hammel, J.~B. Hutchings, P.~Jakobsen, S.~J. Lilly, K.~S. Long, J.~I. Lunine,
  M.~J. Mccaughrean, M.~Mountain, J.~Nella, G.~H. Rieke, M.~J. Rieke, H.-W.
  Rix, E.~P. Smith, G.~Sonneborn, M.~Stiavelli, H.~S. Stockman, R.~A.
  Windhorst, and G.~S. Wright, ``{The James Webb Space Telescope},'' {\em Space
  Sci Rev} {\bf 123}, 485  (2006).

\bibitem{Gennaro:2018}
M.~Gennaro, {\em {WFC3 Data Handbook, Version 4.0}}.
\newblock Baltimore: STScI, 4.0~ed.  (2018).

\bibitem{Vacca:2004fg}
W.~D. Vacca, M.~C. Cushing, and J.~T. Rayner, ``Nonlinearity corrections and
  statistical uncertainties associated with near‐infrared arrays,'' {\em
  PASP} {\bf 116}(818), 352  (2004).

\bibitem{Hilbert:2014uy}
B.~Hilbert, ``{Updated Non-Linearity Calibration Method for WFC/IR},'' {\em
  STScI Instrument Science Report} {\bf 2014-17}, 1--32  (2014).

\bibitem{Piquette:2014cc}
E.~C. Piquette, W.~McLevige, J.~Auyeung, and A.~Wong, ``{Progress in
  development of H4RG-10 infrared focal plane arrays for WFIRST-AFTA},'' in
  {\em Proc SPIE},  A.~D. Holland and J.~Beletic, Eds., 91542H--8, Proc SPIE
  (2014).

\bibitem{Loose:2003vh}
M.~Loose, M.~C. Farris, J.~D. Garnett, D.~N.~B. Hall, and L.~J. Kozlowski,
  ``{HAWAII-2RG: a 2k x 2k CMOS multiplexer for low and high background
  astronomy applications},'' {\em Proc SPIE} {\bf 4850}, 867  (2003).

\bibitem{Zandian:2016cq}
M.~Zandian, M.~Farris, W.~McLevige, D.~Edwall, E.~Arkun, E.~Holland, J.~E.
  Gunn, S.~Smee, D.~N.~B. Hall, K.~W. Hodapp, A.~Shimono, N.~Tamura,
  M.~Carmody, J.~Auyeung, and J.~W. Beletic, ``{Performance of science grade
  HgCdTe H4RG-15 image sensors},'' in {\em Proc SPIE},  A.~D. Holland and
  J.~Beletic, Eds., 99150F--11, SPIE  (2016).

\bibitem{Lotz:2017ic}
J.~M. Lotz, A.~Koekemoer, D.~Coe, N.~Grogin, P.~Capak, J.~Mack, J.~Anderson,
  R.~Avila, E.~A. Barker, D.~Borncamp, G.~Brammer, M.~Durbin, H.~Gunning,
  B.~Hilbert, H.~Jenkner, H.~Khandrika, Z.~Levay, R.~A. Lucas, J.~MacKenty,
  S.~Ogaz, B.~Porterfield, N.~Reid, M.~Robberto, P.~Royle, L.~J. Smith, L.~J.
  Storrie-Lombardi, B.~Sunnquist, J.~Surace, D.~C. Taylor, R.~Williams,
  J.~Bullock, M.~Dickinson, S.~Finkelstein, P.~Natarajan, J.~Richard,
  B.~Robertson, J.~Tumlinson, A.~Zitrin, K.~Flanagan, K.~Sembach, B.~T. Soifer,
  and M.~Mountain, ``{The Frontier Fields: Survey Design and Initial
  Results},'' {\em ApJ} {\bf 837}, 97--24  (2017).

\bibitem{Jolliffe2011}
I.~Jolliffe, {\em Principal Component Analysis}, 1094--1096.
\newblock Springer Berlin Heidelberg, Berlin, Heidelberg  (2011).

\bibitem{Rieke:2007fn}
G.~H. Rieke, ``{Infrared detector arrays for astronomy},'' {\em Annu Rev Astron
  Astr} {\bf 45}, 77--115  (2007).

\bibitem{0471504475}
R.~Courant and D.~Hilbert, {\em Methods of Mathematical Physics, Vol. 1},
  Wiley-VCH  (1989).

\bibitem{Press:1992:NRC:148286}
W.~H. Press, S.~A. Teukolsky, W.~T. Vetterling, and B.~P. Flannery, {\em
  Numerical Recipes in C (2Nd Ed.): The Art of Scientific Computing}, Cambridge
  University Press, New York, NY, USA  (1992).

\bibitem{2005PASP..117...94O}
J.~D. Offenberg, D.~J. Fixsen, and J.~C. Mather, ``{Memory-Efficient
  Up-the-Ramp Processing with Cosmic-Ray Rejection},'' {\em PASP} {\bf 117},
  94--103  (2005).

\bibitem{1991SPIE.1541..127F}
A.~M. Fowler and I.~Gatley, ``{Noise reduction strategy for hybrid IR
  focal-plane arrays},'' {\em Proc SPIE} {\bf 1541}, 127--133  (1991).

\bibitem{Garnett:1993vi}
J.~D. Garnett and W.~J. Forrest, ``{Multiply sampled read-limited and
  background-limited noise performance},'' {\em Proc SPIE} {\bf 1946}, 395
  (1993).

\bibitem{2010PASP..122.1254.}
B.~J. Rauscher, ``Erratum,'' {\em PASP} {\bf 122}, 1254  (2010).

\bibitem{Rauscher:2007gc}
B.~J. Rauscher, O.~Fox, P.~Ferruit, R.~J. Hill, A.~Waczynski, Y.~Wen,
  W.~Xia-Serafino, B.~Mott, D.~Alexander, C.~K. Brambora, R.~Derro, C.~Engler,
  M.~B. Garrison, T.~Johnson, S.~S. Manthripragada, J.~M. Marsh, C.~Marshall,
  R.~J. Martineau, K.~B. Shakoorzadeh, D.~Wilson, W.~D. Roher, M.~Smith,
  J.~Garnett, M.~Loose, S.~Wong-Anglin, M.~Zandian, E.~Cheng, T.~Ellis,
  B.~Howe, M.~Jurado, G.~Lee, J.~Nieznanski, P.~Wallis, J.~York, M.~W. Regan,
  D.~N.~B. Hall, T.~B{\"o}ker, G.~De~Marchi, P.~Jakobsen, and P.~Strada,
  ``{Detectors for the James Webb Space Telescope Near-Infrared Spectrograph.
  I. Readout Mode, Noise Model, and Calibration Considerations},'' {\em PASP}
  {\bf 119}, 768  (2007).

\bibitem{2017PASP..129j5003R}
B.~J. Rauscher, R.~G. Arendt, D.~J. Fixsen, M.~A. Greenhouse, M.~Lander,
  D.~Lindler, M.~Loose, S.~H. Moseley, D.~B. Mott, Y.~Wen, D.~V. Wilson, and
  C.~Xenophontos, ``{Improved Reference Sampling and Subtraction: A Technique
  for Reducing the Read Noise of Near-infrared Detector Systems},'' {\em PASP}
  {\bf 129}, 105003--  (2017).

\bibitem{Anderson:1952fw}
T.~W. Anderson and D.~A. Darling, ``{Asymptotic Theory of Certain Goodness of
  Fit Criteria Based on Stochastic Processes},'' {\em Annals of Mathematical
  Statistics} {\bf 23}(2), 193--212  (1952).

\bibitem{Stephens:1974}
M.~A. Stephens, ``{EDF Statistics for Goodness of Fit and Some Comparisons},''
  {\em Journal of the American Statistical Association} {\bf 69}(347), 730--737
   (1974).

\end{thebibliography}
\bibliographystyle{spiejour}   


\vspace{2ex}\noindent\textbf{Bernard J. Rauscher} is an experimental astrophysicist at NASA's Goddard Space Flight Center. His research interests include astronomy instrumentation, extragalactic astronomy and cosmology, and most recently the search for life on other worlds. Rauscher's work developing detector systems for \JWST was recognized by a (shared) Congressional Space Act award and NASA's Exceptional Achievement Medal.

\vspace{2ex}\noindent\textbf{Richard G. Arendt} is a Senior Research Scientist at UMBC, working in the Observational Cosmology Laboratory at NASA's Goddard Space Flight Center. He has performed software and data analysis tasks related to the {\it COBE}/DIRBE and {\it Spitzer}/IRAC instruments, while researching dust in the solar system and Galaxy, supernova remnants, and the comic infrared background.

\vspace{2ex}\noindent\textbf{Dale Fixsen} has worked at NASA's Goddard Space Flight Center calibrating the FIRAS instrument on the Cosmic Background Explorer (COBE), designing and building the TopHat instrument, and designing the data pipeline for the Infrared Array Camera of the Spitzer Space Telescope. Prior to that, he worked on SQUID magnetic gradiometers for submarine detection and automated data fusion. He holds a PhD in astrophysics from Princeton University and a bachelor's degree in mathematics/physics from Pacific Lutheran University.


\vspace{2ex}\noindent\textbf{Gregory Mosby, Jr.} is a NASA Postdoctoral Program Fellow at NASA Goddard Space Flight Center. He received his BS degree in astronomy and physics from Yale University in 2009, and his MS and PhD degrees in astronomy from the University of Wisconsin - Madison in 2011 and 2016, respectively. His current research interests include near infrared detectors, astronomical instrumentation, and applications of machine learning to observational astronomy. He is a member of SPIE.

\vspace{2ex}\noindent\textbf{Harvey Moseley} is a Senior Astrophysicist at NASA Goddard Space Flight Center. Moseley has received many awards for his work, including a (shared) Gruber Cosmology Prize, the American Astronomical Society's Joseph Weber Award, and SPIE's George W. Goddard Award.

\vspace{1ex}

\listoffigures
\listoftables

\end{spacing}
\end{document}